\documentclass[11pt]{article}

\usepackage{graphicx}                  
\usepackage{dcolumn}                   
\usepackage{bm}

\usepackage{jheppub}                   

\makeatletter
\gdef\@fpheader{\ }                    
\makeatother

\usepackage{graphicx}
\usepackage{dcolumn}
\usepackage{bm}
\usepackage{float}

\usepackage{amsmath,amssymb}           
\usepackage{amscd}                     
\usepackage{epsfig}                    
\usepackage[matrix,arrow]{xy}          
\usepackage{xspace}                    
\usepackage{stmaryrd}                  
\usepackage{slashed}
\usepackage{tikz}
\usepackage{hyperref}
\usepackage[vcentermath]{youngtab}

\newcommand{\bq}{\begin{equation}}
\newcommand{\eq}{\end{equation}}
\newcommand{\bea}{\begin{eqnarray}}
\newcommand{\eea}{\end{eqnarray}}

\newcommand{\dd}{\mathrm{d}}
\newcommand{\ee}{\mathrm{e}}
\newcommand{\ii}{\mathrm{i}}

\newcommand{\der}{\partial}

\newcommand{\bbR}{\mathbb{R}}
\newcommand{\bbC}{\mathbb{C}}

\DeclareMathOperator{\SU}{\mathit{SU}}
\DeclareMathOperator{\SO}{\mathit{SO}}
\DeclareMathOperator{\SL}{\mathit{SL}}

\DeclareMathOperator{\Symp}{\mathit{USp}}
\DeclareMathOperator{\Spin}{\mathit{Spin}}

\newcommand{\rep}[1]{\mathbf{#1}}
\newcommand{\repp}[2]{(\rep{#1}, \rep{#2})}

\newcommand{\Gs}[1]{\Gamma(#1)}

\newcommand{\ph}[1]{\phantom{#1}}
\newcommand{\hs}[1]{\hspace{#1}}
\newcommand{\ra}{\rightarrow}
\newcommand{\Ra}{\Rightarrow}

\newcommand{\Lgen}{L}
\newcommand{\LgenS}{L^{{}^{\text{KD}}}}

\newcommand{\Dgen}{{D}}

\DeclareMathOperator{\adj}{ad}

\newcommand{\LC}{\nabla}

\newcommand{\proj}[1]{\times_{#1}}

\DeclareMathOperator{\Edd}{\mathit{E_{d(d)}}}

\DeclareMathOperator{\dHd}{\mathit{\tilde{H}_d}}

\newcommand{\tA}{{\tilde{A}}}
\newcommand{\tF}{{\tilde{F}}}

\newcommand{\tV}{\tilde{V}}

\newcommand{\am}{Q}

\newcommand{\Kgen}{K}
\newcommand{\HConSp}{\Kgen_{\SU(8)}}

\newcommand{\Tgen}{W}		 	
\newcommand{\Tint}{\Tgen_{\text{int}}}

\newcommand{\CC}{\text{c.c.}}

\newcommand{\gN}{G_{\mathcal{N}}}

\newcommand{\Xa}{(X_1)}
\newcommand{\Xb}{(X_2)}
\newcommand{\Xc}{(X_3)}
\newcommand{\Xd}{(X_4)}
\newcommand{\Xe}{(X_5)}
\newcommand{\Xf}{(X_6)}

\newcommand{\Ya}{(Y_1)}
\newcommand{\Yb}{(Y_2)}
\newcommand{\Yc}{(Y_3)}

\newcommand{\gamh}{\hat{\gamma}}

\title{Supersymmetric Backgrounds, the Killing Superalgebra, and  Generalised Special Holonomy}

\author[a]{Andr\'e Coimbra,}
\emailAdd{coimbra@ihes.fr}
\author[a,b]{Charles Strickland-Constable}
\emailAdd{charles.strickland-constable@cea.fr}

\affiliation[a]{Institut des Hautes \'Etudes Scientifiques, Le Bois-Marie, 35 route de Chartres, F-91440 Bures-sur-Yvette, France}

\affiliation[b]{Institut de physique th\'eorique, 
	Universit\'e Paris Saclay, CEA, CNRS, F-91191 Gif-sur-Yvette, France}

\subheader{\textrm{IPhT-T16/064}
}

\abstract{We prove that, for M theory or type II, generic Minkowski flux backgrounds preserving $\mathcal{N}$ supersymmetries in dimensions $D\geq4$ correspond precisely to integrable generalised $G_{\mathcal{N}}$ structures, where $G_{\mathcal{N}}$ is the generalised structure group defined by the Killing spinors. In other words, they are the analogues of special holonomy manifolds in $E_{d(d)} \times\mathbb{R}^+$ generalised geometry. In establishing this result, we introduce the Kosmann-Dorfman bracket, a generalisation of Kosmann's Lie derivative of spinors. This allows us to write down the internal sector of the Killing superalgebra, which takes a rather simple form and whose closure is the key step in proving the main result. In addition, we find that the eleven-dimensional Killing superalgebra of these backgrounds is necessarily the supertranslational part of the $\mathcal{N}$-extended super-Poincar\'e algebra.
\vfill}

\begin{document}  
 
\maketitle





\section{Introduction}
\label{sec:intro}

Generalised geometry, first defined by Hitchin and Gualtieri~\cite{Hitchin:2004ut,Gualtieri:2003dx}, has given physicists the tools to understand in a fully geometric formalism the symmetries of the abelian $p$-form gauge fields of low energy string theory.
Sections of the generalised tangent bundle, which is isomorphic to the sum of the tangent and cotangent bundle, generate not only the diffeomorphism symmetry of usual Riemannian geometry, but also the gauge symmetry of the $B$ field, which is reinterpreted as a connection on a gerbe that describes the twisting of the cotangent over the tangent bundle. The topological data of both the manifold and the flux $H = \dd B$ are thus also encoded.
This generalised tangent bundle, which possesses a natural $O(d,d)$ structure, is the correct setting for understanding the NSNS sector of type II supergravity (allowing for some small modifications to incorporate the dilaton~\cite{CSW1}). In order to geometrise the remaining RR fields, or alternatively M theory, exceptional generalised geometry was developed in~\cite{chris,PW,CSW2,CSW3}, which expands the generalised tangent bundle to admit an $\Edd\times\bbR^+$ structure, corresponding to the U-duality groups. Further versions of generalised geometry, relevant for other supergravities, have since been introduced~\cite{Baraglia:2011dg,Bouwknegt,Hitchin-Bn,Rubio:2013ada,Strickland-Constable:2013xta,LSW1,Garcia-Fernandez:2013gja,CMTW,Cassani:2016ncu}.

An obvious field to apply these new ideas is the problem of obtaining solutions for generic compactifications of string and M theory with fluxes, since the gauge fields are now deeply integrated in the formalism. Of particular interest are those that preserve supersymmetry. Supersymmetric backgrounds are often thought of in the language of $G$-structures~\cite{Gauntlett:2002sc,Gauntlett:2002fz,Martelli:2003ki}. In the absence of fluxes, preserving supersymmetry implies the existence of global spinors which are parallel with respect to the Levi-Civita connection, or put in another way, the Killing spinors define a $G$-structure which is torsion-free. As is well known, this is a very constraining condition, since it immediately restrict solutions to be special holonomy manifolds. The addition of fluxes breaks this integrability condition however, and instead one is lead to attempt to classify solutions based on how the fluxes arrange themselves into the  intrinsic torsion classes of the $G$-structures defined by the Killing spinors. One can go even further, and consider \textit{generalised} $G$-structures~\cite{GMPT}, that is $G$-structures on the generalised tangent space, which naturally provide a unified description of flux background~\cite{GMPT,PW,GO} and so allow more extensive and systematic analyses of solutions (see, for example,~\cite{Grana:2006kf,Andriot:2008va,Minasian:2006hv,Butti:2007aq,Gabella:2009wu,GCG-AdS}).

In~\cite{CSW4} a notion of integrability for generalised $G$-structures was developed based on the Dorfman bracket of generalised vector fields which is suitable for any version generalised geometry. In particular, when applied to exceptional generalised geometry, it was proven that the condition of vanishing generalised intrinsic torsion of the $G$-structure defined by a global spinor corresponds precisely to Minkowski flux backgrounds of M theory or type II preserving $\mathcal{N}=1$. In other words, it became possible to describe fully generic backgrounds as precisely the generalised analogue of special holonomy manifolds -- for example, a torsion-free generalised $SU(7)$ structure on a 7-fold is the full flux generalisation of the fluxless $G_2$ manifold solution for M theory compactifications to four-dimensional flat spacetime. Generalised torsion-free spaces were therefore dubbed ``generalised special holonomy spaces''.\footnote{Note that the term generalised special holonomy is used strictly in analogy to torsion-free $G$-structures defined by spinors. We do not define any new notion of holonomy as a arising from some generalisation of parallel transport. A different notion of ``generalised holonomy'' had also been defined in~\cite{Duff:2003ec,Hull:2003mf}, referring instead to the holonomy of the (non-generalised) connection appearing in the gravitino variation.}

This machinery has since been used to describe AdS backgrounds that preserve minimal supersymmetry in~\cite{CS1}, which turn out to precisely correspond to spaces with constant singlet torsion, i.e. ``weak generalised special holonomy'' spaces, and in~\cite{Ashmore:2015joa} the generalised geometry integrability conditions for 8 supercharge vacua were shown to be rephrasable in terms of hyper- (H) and vector- (V) multiplet structures on the generalised tangent space following the earlier work~\cite{GLSW}. This allowed the proof of some basic features of the moduli space of such generalised structures (and thus of flux vacua), such as the K\"ahler and hyper-K\"ahler structures necessitated by having 8 supercharges in the external space theory. 
This construction was then applied to AdS backgrounds and holography~\cite{Ashmore:2016qvs} showing how features of the dual field theory such as the central charge and volume minimisation could be encoded in the generalised geometry. The integrability of the H- and V-structures was also explicitly shown to follow from the Killing spinor equations for AdS$_5$ in~\cite{Grana:2016dyl}.
Subsequently, a holographic description of the marginal deformations of four-dimensional $\mathcal{N}=1$ SCFTs was given in~\cite{Ashmore:2016oug}, where in particular, it was shown that in general the exactly marginal deformations correspond to a quotient of the classically marginal deformations by the action of the isometry group of the internal space, reproducing an earlier field-theoretical result~\cite{Green:2010da}. These works provide a good demonstration of the power of using a generalised formalism which treats fluxes and geometry on equal footing.

However, the analysis of~\cite{CSW4} was, in a sense, incomplete, since the proof of the correspondence between supersymmetric solutions and generalised special holonomy spaces was only given for the $\mathcal{N}=1$ case. While it was shown that for each value of $\mathcal{N}$ there exists a single unified $\gN$ structure on the generalised tangent bundle (an extension of the argument first presented in~\cite{PW}), that its generalised intrinsic torsion vanishes for supersymmetric Minkowski solutions was only proven when there is a single Killing spinor. The $\mathcal{N}=2$ case could actually be derived in the exact same manner, as was shown explicitly in~\cite{Ashmore:2015joa}, but if one were to attempt to reproduce the same steps for $\mathcal{N}\geq 3$ there would appear to exist a mismatch between a naive counting of the representations appearing in the generalised intrinsic torsion space of the $\gN$ structure and the constraints provided by the $\mathcal{N}$ Killing spinor equations.

In this paper we show how this issue is resolved and prove that, indeed, all supersymmetric Minkowski flux backgrounds of M theory and type II are in one-to-one correspondence with generalised special holonomy manifolds. For each solution with an amount $\mathcal{N}$ of preserved supersymmetry there exists a single corresponding torsion-free generalised $\gN$ structure and vice versa.

Intuitively, the key observation is that the supergravity fluxes do not fill out the entirety of the generalised torsion $\gN$ representations. This can be traced back to the fact that the torsion is fixed by the Dorfman bracket which, since it involves the anchor map explicitly, acts only within what is often called the ``geometric subgroup'' -- that is the parabolic subgroup induced by the supergravity bosonic symmetries, corresponding to infinitesimal diffeomorphisms and gauge transformations -- and not the entire $\Edd\times\bbR^+$ generalised frame group. 

In order to exploit these properties explicitly at the level of the $\gN$ structure defined by the Killing spinors, we introduce in section~\ref{sec:gen-geom} an extension of the Dorfman bracket that can act on spinors, the Kosmann-Dorfman bracket. This is the direct analogue of Kosmann's generalisation of the Lie derivative for spinor fields~\cite{Kosmann}. With this new tool we are able to explore in generalised geometry the algebra generated by the internal Killing spinors. This algebra is essentially part of the eleven-dimensional ``Killing superalgebra". 

In (pseudo-)Riemannian geometry it is natural to consider the Lie algebra of the Killing vector fields of a metric, which forms the Lie algebra of the isometry group of the manifold. When one considers supergravity solutions, one would like to consider isometries of the full background, i.e. vectors which preserve all of the fields rather than just the metric. In addition, one could also consider the Killing spinors, which generate the (infinitesimal) supersymmetry transformations that leave the background invariant. In eleven-dimensional supergravity, it was shown in~\cite{FigueroaO'Farrill:2004mx} by checking their closure and  Jacobi identities, that all of these transformations naturally form a superalgebra, called the Killing superalgebra, generated by both the Killing vectors and Killing spinors of the background. Similar results were also found in ten-dimensional supergravities~\cite{FigueroaO'Farrill:2007ar}. 
In these works, it was shown that any background preserving more that 24 supercharges (or 16 supercharges in the heterotic case) must be locally homogeneous.
Later, this was extended to a proof of the homogeneity theorem~\cite{FigueroaO'Farrill:2012fp}, which says that any supersymmetric background with more than 16 preserved supercharges must be locally homogeneous. Further, there is a spanning set of Killing vectors which arises from the bilinears of the Killing spinors. The Killing superalgebras and corresponding homogeneity results for six-dimensional $(1,0)$ and $(2,0)$ supergravity were also given in~\cite{Figueroa-O'Farrill:2013aca}. The homogeneity results are useful as it is possible to classify the homogeneous solutions of the equations of motion~\cite{FigueroaO'Farrill:2011fj,FigueroaO'Farrill:2012rf}.

In section~\ref{sec:KSA} we will find that the internal part of the Killing superalgebra has a neat manifestation in the language of $\Edd\times\bbR^+$ generalised geometry. 
It is already known that isometries which preserve the supergravity fields are precisely   those generated by generalised Killing vectors~\cite{GMPW}, i.e. generalised vectors such that the Dorfman derivative of the generalised metric along them vanishes. In fact, the parabolic nature of the Dorfman derivative implies that when it is evaluated along a generalised Killing vector in a frame ``untwisted'' by the gauge fields, it reduces to an ordinary Lie derivative along a genuine Killing vector field, as was also noted in~\cite{Ashmore:2015joa,Grana:2016dyl}. We find that this applies to the Kosmann-Dorfman derivative of spinors as well and give it a concrete proof in appendix~\ref{app:spinor-Dorfman-GKV}. This lemma turns out to great simplify otherwise cumbersome computations, and we are able to explore the algebra generated by the Killing spinors in terms of this new bracket.
Further, we have that Killing spinor bilinears on the internal space give rise to generalised Killing vectors. Therefore this formulation  automatically encodes the $p$-form gauge transformations generated by the brackets of supersymmetries as well as the diffeomorphisms described by the usual Killing superalgebra, as the generalised vectors also include differential form components which are precisely the generators of the gauge transformations of the theory. The Kosmann-Dorfman derivative then includes a natural way to define the action of these form bilinears on the Killing spinors (the problem of finding such an action was previously raised in~\cite{FigueroaO'Farrill:2008ka,Ertem:2016fso}).
We find that the Killing spinors generate an algebra which is simpler than one might naively expect -- in fact, we prove that all brackets other than the spinor bilinear vanish for external Minkowski solutions. As a corollary, we give an eleven-dimensional interpretation of this result, showing that it implies that the eleven-dimensional Killing spinors generate the supertranslational ideal of the $\mathcal{N}$-extended super-Poincar\'e algebra of the external Minkowski space.

In section~\ref{sec:gen-hol}, not only do we reproduce the result of~\cite{CSW4}, we find that the components of the generalised intrinsic torsion which initially appeared to be unconstrained by the Killing spinor equations can be rewritten in terms of this new spinorial bracket, and in fact turn out to be precisely just one of the trivial structure constants of the subalgebra generated by the Killing spinors. We can then simply conclude that these last remaining components of the torsion vanish as well. The generalised $\gN$ structures describing $\mathcal{N}$ supersymmetric backgrounds are indeed exactly those which are torsion-free.

Note that the same caveats as in~\cite{CSW4} apply to our analysis here. We do not address the no-go theorems for Minkowski flux compactifications~\cite{Candelas:1984yd,Maldacena:2000mw,Ivanov:2000fg}, so the spaces we study (we only examine local structures) should be taken to either be non-compact or with boundaries. Also, for a majority of this paper we will be working on the specific case of $d=7$ in the context of M theory compactifications. This corresponds to an external four-dimensional Minkowski space and an internal manifold whose description is governed by $E_{7(7)}\times\bbR^+$ generalised geometry. This is the largest of the known exceptional generalised geometries, and using the constructions of~\cite{CSW2,CSW3} it is possible to obtain the $d<7$ formulations from it via straightforward truncations and decompositions. Alternatively, one can follow the procedures outlined in~\cite{CSW2} to find the $d-1$ internal geometry of a Type II compactification in a democratic formalism. The $d>7$ cases are, for the moment, beyond our scope. On the other hand, one should note that in $d=4$ the generalised holonomy groups already coincide with usual geometrical ones, so we skip the discussion of $d\leq3$ as the generalised geometry description is unnecessary.



\section{Supergravity preliminaries}


\subsection{The Killing superalgebra in eleven-dimensions}

We begin by briefly recounting the construction of the Killing superalgebra in eleven-dimensions, following~\cite{FigueroaO'Farrill:2004mx}. Here, we are working on a solution $(\mathcal{M},\mathcal{G},\mathcal{F})$ of eleven-dimensional supergravity, where $\mathcal{M}$ is a smooth spin manifold with metric $\mathcal{G}$ of mostly-plus signature $(10,1)$ and $\mathcal{F}$ is the four-form flux. All fermions are set to zero.

Let $\mathfrak{g}_0$ be the vector space of Killing vectors of $(\mathcal{M},\mathcal{G},\mathcal{F})$, i.e. vectors $v$ such that $\mathcal{L}_v\mathcal{G}=\mathcal{L}_v\mathcal{F}=0$, and $\mathfrak{g}_1$ be the space of Killing spinors, i.e. (commuting) spinor fields $\varepsilon$ satisfying
\begin{equation}
\label{eq:11d-susy}
	\delta_\varepsilon \Psi_M = \LC_M \varepsilon
		+ \tfrac{1}{288} (\Gamma_M{}^{P_1 \dots P_4} \mathcal{F}_{P_1 \dots P_4} 
			- 8 \mathcal{F}_{MN_1 \dots N_3} \Gamma^{N_1 \dots N_3}) \varepsilon = 0 .
\end{equation}
Then the central point is that, for $v, v_1, v_2 \in \mathfrak{g}_0$ and $\varepsilon, \varepsilon_1, \varepsilon_2 \in \mathfrak{g}_1$ the bracket operations
\begin{equation}
\label{eq:KSA}
\begin{aligned}[]
	[ \varepsilon_1, \varepsilon_2 \} &= v(\varepsilon_1, \varepsilon_2) ,\\
	[ v , \varepsilon \} &= \mathcal{L}_v \varepsilon ,\\
	[v_1, v_2\}  &= \mathcal{L}_{v_1} v_2 ,
\end{aligned}
\end{equation}
where $v(\varepsilon_1, \varepsilon_2)^M = \bar\varepsilon_1 \Gamma^M \varepsilon_2$ and $\mathcal{L}_v\varepsilon$ is the Kosmann derivative of a spinor field~\cite{Kosmann}, make $\mathfrak{g} = \mathfrak{g}_0 \oplus \mathfrak{g}_1$ into a Lie superalgebra. This is the Killing superalgebra. Checking that~\eqref{eq:KSA} satisfies the Jacobi identities necessary for the definition of a Lie superalgebra is a non-trivial task; we refer the reader to~\cite{FigueroaO'Farrill:2004mx} for full details. One can also make the analogous construction for type II backgrounds and find that there exists a Killing superalgebra structure there as well~\cite{FigueroaO'Farrill:2007ar}.


\subsection{Internal sector of dimensional split}

Turning now to the problem of characterising $D$-dimensional Minkowski backgrounds of M theory, where $D = 11 - d$, we will very briefly recall the supergravity setup in the specific case $d=7$, for which the internal space is described by $E_{7(7)}\times\bbR^+$ generalised geometry. For the supergravity ansatze with $d < 7$ (as well as a more detailed treatment of $d=7$) see~\cite{CSW3}.

We consider a dimensional split $11 \ra (3,1)+7$, and concretely a warped product metric
\begin{equation}
\label{eq:warped-metric}
	\dd s^2_{10,1} = \ee^{2\Delta} \eta_{\mu \nu} \dd x^\mu \dd x^\nu + g_{mn} \dd y^m \dd y^n ,
\end{equation}
where the first term corresponds to an external four-dimensional Minkowski space, and the second one to an internal curved space. All fields are taken to depend only on the internal coordinates, and we allow only the internal fluxes $F_{mnpq} = \mathcal{F}_{mnpq}$ and $\tF_{m_1 \dots m_7} = (*_{11} \mathcal{F})_{m_1 \dots m_7}$. 

As such, we can decompose the eleven-dimensional spinors into products of external and internal spinors,\footnote{
Note that the possibility of more general Killing spinors where there is linear dependence on the external coordinates is raised in~\cite{Gutowski:2014ova,Beck:2014zda,Beck:2015hpa}, and explicit examples appear in~\cite{Papadopoulos-future}.
However, we restrict attention to the ansatz~\eqref{eq:spinor-decomp}, which we view as giving an uplift of the usual external Killing spinors to higher-dimensions, and this is what we will refer to as a generic supersymmetric background in this paper.
} following exactly appendix (C.4) of~\cite{CSW3}. We define seven-dimensional (commuting) Killing spinors $\hat\epsilon$ to be non-vanishing complex spinors satisfying the internal Killing spinor equations
\begin{equation}
\label{eq:7d-susy}
\begin{split}
	\mathcal{D}_m\hat\epsilon_i := \LC_m \hat\epsilon_i + \tfrac{1}{288} 
		(\gamma_m{}^{n_1 \dots n_4} - 8 \delta_{m}{}^{n_1} \gamma^{n_2 n_3 n_4}) 
   		F_{n_1 \dots n_4}  \hat\epsilon_i
   		- \tfrac{1}{12} \tfrac{1}{6!} \tilde{F}_{mn_1 \dots n_6}
			\gamma^{n_1 \dots n_6} \hat\epsilon  &=0,\\[8pt]
	 \mathcal{D}\hat\epsilon_i:=\gamma^m\LC_m  \hat\epsilon_i 
		+ \gamma^m(\der_m \Delta)  \hat\epsilon_i 
			- \tfrac{1}{96} \gamma^{m_1 \dots m_4} F_{m_1 \dots m_4}  \hat\epsilon_i 
				- \tfrac{1}{4} \tfrac{1}{7!} \gamma^{m_1 \dots m_7} 
					\tilde{F}_{m_1 \dots m_7}  \hat\epsilon  &=0 .
\end{split}
\end{equation}
Given an internal Killing spinor $\hat\epsilon$ solving~\eqref{eq:7d-susy} and a constant four-dimensional Majorana spinor $\eta$, we have that
\begin{equation}
\label{eq:spinor-decomp}
	\varepsilon = \eta^+ \otimes \hat\epsilon + \eta^- \otimes (\tilde{D} \hat\epsilon)^*,
	\hs{40pt}
	\gamma^{(4)} \eta_{\pm} = \mp \ii \eta_{\pm} ,
\end{equation}
is an eleven-dimensional Killing spinor solving~\eqref{eq:11d-susy} (see e.g.~\cite{CSW4}), motivating the above definition. We say that an $\mathcal{N}$ supersymmetric Minkowski background is one equipped with $\mathcal{N}$ independent Killing spinors as defined here (so that the space of Killing spinors is an $\mathcal{N}$ dimensional complex vector space). For each internal Killing spinor, the formula~\eqref{eq:spinor-decomp} then gives an uplift of any constant external Majorana spinor, that is a Killing spinor of Minkowski space, to a higher-dimensional Killing spinor.

Crucially for what follows, one can take the internal Killing spinors to be orthonormal~\cite{Gabella:2012rc}. We provide a slightly different proof which generalises more readily to other dimensions and to type II. As in~\cite{Gabella:2012rc}, it will be very useful to rewrite the system~\eqref{eq:7d-susy} in the equivalent form
\begin{equation}
\label{eq:7d-susy-mod}
\begin{aligned}
	\nabla_m \hat\epsilon - \tfrac12 (\der_m \Delta) \hat\epsilon 
		- \tfrac12 (\der_n \Delta) \gamma_m{}^n \hat\epsilon
		- \tfrac14 \tfrac{1}{3!} F_{mnpq} \gamma^{npq} \hat\epsilon
		- \tfrac14 \tfrac{1}{6!} \tF_{mn_1 \dots n_6} \gamma^{n_1 \dots n_6} \hat\epsilon &= 0, \\
		\tfrac{1}{12} \tfrac{1}{4!}F_{m_1 \dots m_4}\gamma^{m_1 \dots m_4}\epsilon + \tfrac16 \tfrac{1}{7!}\tF_{m_1 \dots m_7}\gamma^{m_1 \dots m_7} \hat\epsilon 
		+ \tfrac12 (\der_m \Delta) \gamma^m\hat\epsilon &= 0.
\end{aligned}
\end{equation}
Defining the usual rescaled supersymmetry generator (c.f.~\cite{CSW3,Gabella:2012rc})
\begin{equation}
\label{eq:rescaled-spinor}
	\epsilon = \ee^{-\Delta/2} \hat\epsilon ,
\end{equation}
we observe that the first of equations~\eqref{eq:7d-susy-mod} has the form
\begin{equation}
\label{eq:der-SU8}
	\tilde\LC_m \epsilon = \nabla_m \epsilon
		- \tfrac14 \tfrac{1}{3!} F_{mnpq} \gamma^{npq} \epsilon
		- \tfrac14 \tfrac{1}{6!} \tF_{mn_1 \dots n_6} \gamma^{n_1 \dots n_6} \epsilon = 0 ,
\end{equation}
where $\tilde\LC$ is an $\SU(8)$ connection. This provides the extremely useful result that given any two Killing spinors $\epsilon_i$ and $\epsilon_j$, the rescaled complex inner product is constant over the seven-dimensional space
\begin{equation}
	\der_m ({\epsilon_i}{}^\dagger \epsilon_j)
	= \tilde\LC_m ({\epsilon_i}{}^\dagger \epsilon_j) = 0 .
\end{equation}
Therefore, we can always find a basis for the space of Killing spinor fields which is orthonormal, i.e. 
\begin{equation}
\label{eq:orth-sp}
	{\epsilon_i}{}^\dagger \epsilon_j = \delta^i{}_j .
\end{equation}
The exact same will hold for type II, and for $d<7$ one just obtains the relevant truncated version of~\eqref{eq:der-SU8}, which will give a connection on a smaller, but still norm-preserving, group (precisely the same local groups we encounter in generalised geometry), so this proof works universally for Minkowski backgrounds.

Finally, we can use an identical argument to that presented in~\cite{FigueroaO'Farrill:2004mx} to deduce that, for any spinor $\zeta$ and Killing vector $v$, the supersymmetry operators~\eqref{eq:7d-susy} obey 
\begin{equation}
\label{eq:comm-susy-lie}
\begin{aligned}
\mathcal{L}_v \left( \mathcal{D}_m\zeta\right) &=\mathcal{D}_m\left(\mathcal{L}_v \zeta\right)  ,\\
\mathcal{L}_v \left(\mathcal{D}\zeta\right) &=\mathcal{D}\left(\mathcal{L}_v \zeta\right) .
\end{aligned}
\end{equation}
In particular, given a Killing spinor $\epsilon$ and a Killing vector of the flux background $v$, we have that the Lie derivative $\mathcal{L}_v \epsilon$ is another Killing spinor. This means that given the above basis of Killing spinors we can introduce constant coefficients $X_i{}^j$ associated to a given Killing vector $v$ such that
\begin{equation}
\label{eq:Lie-KS}
	\mathcal{L}_v \epsilon_i = X_i{}^j \epsilon_j .
\end{equation}
As $i_v \dd \Delta = 0$, we can make an identical statement for the rescaled spinors $\hat\epsilon_i$. Thus, the Killing spinors form a representation of the isometry algebra of the background.


\section{Generalised geometry preliminaries}
\label{sec:gen-geom}

In this section we will introduce some additional tools from generalised geometry that were not covered in~\cite{CSW4} but which we will need to establish our results. For an introduction to $\Edd\times\bbR^+$ generalised geometry and the notation that will follow, please see~\cite{CSW2,CSW3}.

\subsection{Generalised Killing vectors}

\begin{table}[htb]
\centering
\begin{tabular}{llll}
   $d$ & $\dHd$ & $S$ & $J$  \\[3pt]
   \hline \\[-12pt]
   7 & $\SU(8)$ & $\rep{8}$ & $\rep{56}$ 
    \\[3pt]
   6 & $ \Symp(8)$ & $\rep{8}$ & $\rep{48}$  \\[3pt]
   5 & $\Symp(4)\times\Symp(4)$ & $\repp{4}{1} + \repp{1}{4}$ 
      & $\repp{4}{5} + \repp{5}{4}$  \\[3pt]
   4 & $\Symp(4)$ & $\rep{4}$ & $\rep{16}$	
\end{tabular}
\caption{Double covers of the maximal compact subgroups of $\Edd\times\bbR^+$ and the representations of the bundles $S$ and $J$ corresponding to spinors and vector-spinors respectively in $\Spin(d)$. Note that $\Symp(2n)$ denotes the compact symplectic group of rank $n$.}
\label{table}
\end{table}
Recall that the supergravity fields $(g, A, \tA, \Delta)$ form a generalised metric $G$, that is a positive-definite inner-product on the $\Edd\times\bbR^+$ generalised tangent bundle $E$. We thus have a reduction of the structure group to $\dHd$, the maximal compact subgroup of $\Edd$ (or rather its double-cover since we will assume our base manifold $M$ is spin), see table~\ref{table}. This means that from the generalised frame bundle $\tilde{F}$ for $E$, which is an $\Edd\times\bbR^+$ principal bundle, we can pick out a subbundle $P \subset \tilde{F}$, the $\dHd$ bundle corresponding to the generalised vielbeins for the metric $G$. We have also that the $\Edd\times\bbR^+$ adjoint bundle $\adj{\tilde{F}} \subset E \otimes E^*$ may be decomposed orthogonally
\begin{equation}
\label{eq:adj-decomp}
	\adj{\tilde{F}} = \adj{P} \oplus \adj{P}^\perp ,
\end{equation}
such that $\adj{P}^\perp$ is also an $\dHd$ bundle.

Following~\cite{GMPW}, a generalised vector field $V \in \Gs{E}$ is called a generalised Killing vector (GKV) if it preserves the generalised metric
\begin{equation}
\label{eq:GKV}
	\Lgen_V G = 0 ,
\end{equation}
where $\Lgen_V$ is the Dorfman derivative, also known as the generalised Lie derivative, along $V$. This definition was introduced in~\cite{GMPW} in the context of $O(d,d)$ generalised geometry, but has subsequently been used for other cases in e.g.~\cite{LSW2,Ashmore:2015joa,Ashmore:2016qvs}. Physically, it means that the generalised vector generates an infinitesimal generalised diffeomorphism (that is, a combined diffeomorphism and gauge transformation) which leaves the background invariant. Writing the Dorfman derivative in the form~\cite{CSW2}
\begin{equation}
	\Lgen_V G = \der_V G - (\der \proj{\adj{\tilde {F}}} V) \cdot G ,
\end{equation}
where  $\proj{}$ denotes the projection of the tensor product to the indicated subspace, it immediately follows for a GKV $V$ 
\begin{equation}
\label{eq:LVG}
	\Lgen_V G = \Lgen^{(D)}_V G = \Dgen_V G - (\Dgen \proj{\adj{\tilde {F}}} V) \cdot G 
		= - (\Dgen \proj{\adj{P}^\perp} V) \cdot G = 0 .
\end{equation}
Here $\Dgen$ is any torsion-free $\dHd$ compatible generalised connection, i.e. a generalised Levi--Civita connection. Thus we deduce that~\eqref{eq:GKV} is equivalent to\footnote{
One might be concerned that there is an ambiguity in the LHS since there exists a family of generalised Levi--Civita connections for a given generalised metric. However,~\eqref{eq:LVG} makes it clear that it is independent of the particular choice of $\Dgen$.}
\begin{equation}
\label{eq:GKV-ad-proj}
	\Dgen \proj{\adj{P}^\perp} V = 0 .
\end{equation}
This is the generalised geometry analogue of the familiar statement that $\LC_{(m} v_{n)} = 0$ for an ordinary Killing vector.

As in~\cite{GMPW}, and focusing on the $d=7$ case for concreteness where
\begin{equation}
\begin{aligned}
	E &\cong TM \oplus \Lambda^2 T^*M \oplus \Lambda^5 T^*M 
		\oplus (T^*M \otimes \Lambda^7 T^*M) ,\\
	V &= v + \omega + \sigma + \tau ,
\end{aligned}
\end{equation}
one can easily find that the condition for $V$ to be a GKV can be written in a coordinate basis as
\begin{equation}
	\mathcal{L}_v g = \mathcal{L}_v \Delta = 0,
	\hs{30pt}
	\mathcal{L}_v A = \dd \omega,
	\hs{30pt}
	\mathcal{L}_v \tA =  \dd \sigma + \tfrac12 A \wedge \dd \omega,
\end{equation}
the last two conditions implying that the fluxes are preserved $\mathcal{L}_v F = \mathcal{L}_v \tF = 0$. Thus, the vector component $v \in \Gs{TM}$ is a Killing vector of the background in the sense of~\cite{FigueroaO'Farrill:2004mx}, i.e. it is a Killing vector which in addition preserves the fluxes.

More useful for our purposes will be the corresponding statement for the components written in what was called in~\cite{CSW2} a \emph{non-conformal} split frame.

One way to think of such frames is in contrast with the \emph{conformal} split frames. The $\tilde{H}_7 = \SU(8)$ generalised metric defines a set of special frames for $E$, the analogue of orthogonal frames in Riemannian geometry. A special class of these are the conformal split frames, which are explicitly constructed in terms of the supergravity fields, and have the generic form
\begin{equation}
\label{eq:conf-split}
\begin{aligned}
   \hat{E}_a &= \ee^{\Delta}\Big( \hat{e}_a + i_{\hat{e}_a} A
      + i_{\hat{e}_a}\tA 
      + \tfrac{1}{2}A\wedge i_{\hat{e}_a}A 
      \\ & \qquad \qquad 
      + jA\wedge i_{\hat{e}_a}\tA 
      + \tfrac{1}{6}jA\wedge A \wedge i_{\hat{e}_a}A \Big) , \\
   \hat{E}^{ab} &=  \ee^{\Delta}\left( e^{ab} + A\wedge e^{ab} 
      - j\tA\wedge e^{ab}
      + \tfrac{1}{2}jA\wedge A \wedge e^{ab} \right) , \\
   \hat{E}^{a_1\dots a_5} &= \ee^{\Delta}\left( e^{a_1\dots a_5} 
      + jA\wedge e^{a_1\dots a_5} \right) , \\
   \hat{E}^{a,a_1\dots a_7} &= \ee^{\Delta}e^{a,a_1\dots a_7} ,
\end{aligned}
\end{equation}
where the $\hat{e}_a$ are orthonormal frames for $TM$ with respect to the Riemannian metric $g$ and $e^a$ the dual frames for $T^*M$. Much as Lorentzian frames are used to introduce fermions in General Relativity, $\SU(8)$ frames allow us to use $\SU(8)$ spinors in $E_{7(7)}\times\bbR^+$ generalised geometry. The conformal split frames are the $\SU(8)$ frames which we can most easily relate to the usual supergravity objects and so most equations we will be writing in the following are naturally expressed in these frames. 

However, note that the ``vector'' component of a generalised vector in such a frame does not coincide the actual vectors we usually deal with in supergravity, as it comes multiplied with the warp factor (this is also the reason why the rescaled spinor in~\eqref{eq:rescaled-spinor} is the natural object in generalised geometry). To obtain genuine vectors, one should instead work on a non-conformal split frame, related to the previous ones by an $\bbR^+$ transformation and taking the form
\begin{equation}
\label{eq:non-conf-split}
\begin{aligned}
   \hat{E}'_a &= \Big( \hat{e}_a + i_{\hat{e}_a} A
      + i_{\hat{e}_a}\tA 
      + \tfrac{1}{2}A\wedge i_{\hat{e}_a}A 
      \\ & \qquad \qquad 
      + jA\wedge i_{\hat{e}_a}\tA 
      + \tfrac{1}{6}jA\wedge A \wedge i_{\hat{e}_a}A \Big) , \\
   \hat{E}'^{ab} &=  \left( e^{ab} + A\wedge e^{ab} 
      - j\tA\wedge e^{ab}
      + \tfrac{1}{2}jA\wedge A \wedge e^{ab} \right) , \\
   \hat{E}'^{a_1\dots a_5} &= \left( e^{a_1\dots a_5} 
      + jA\wedge e^{a_1\dots a_5} \right) , \\
   \hat{E}'^{a,a_1\dots a_7} &= e^{a,a_1\dots a_7} . 
\end{aligned}
\end{equation}
These are, in a sense, frames for a ``conformally-rescaled'' generalised metric, since they satisfy $G(\hat{E}'_A,\hat{E}'_B)=\ee^{-2\Delta}\delta_{AB}$. In such a frame one finds that a generalised Killing vector obeys the relations
\begin{equation}
\label{eq:GKV-non-conf}
	\mathcal{L}_v g = \mathcal{L}_v \Delta = 0,
	\hs{30pt}
	\dd \omega = i_v F,
	\hs{30pt}
	\dd \sigma = i_v \tF - \omega \wedge F ,
\end{equation}
from which we can calculate the components of the Dorfman derivative and obtain the elegant result (this has also recently been noted in~\cite{Ashmore:2015joa,Grana:2016dyl})
\begin{equation}
\label{eq:Dorfman-GKV-Lv}
	\Lgen_V V' = \mathcal{L}_v V' .
\end{equation}
The Dorfman derivative by a GKV thus reduces to the ordinary Lie derivative in the non-conformal split frame.
The conditions~\eqref{eq:GKV-non-conf} on the GKV are exactly such that the additional parts of the Dorfman derivative are cancelled by the flux terms which arise from the twisting of the generalised tangent space.

\subsection{The Kosmann-Dorfman derivative}
In~\cite{Kosmann} Kosmann introduced a notion of Lie derivative of a spinor by a general vector field
\begin{equation}\label{eq:Lie-Spin}
	\mathcal{L}^{{}^\text{K}}_v \epsilon = \LC_v \epsilon + \tfrac14 (\LC_{[a} v_{b]}) \gamma^{ab} \epsilon .
\end{equation}
There exists a straightforward extension of this definition for generalised geometry. As usual~\cite{CSW3}, we now think of $\Spin(d)$ spinors as sections of the $\dHd$ bundle $S$ (see table~\ref{table}). We define the Kosmann-Dorfman (KD) derivative of a spinor $\epsilon\in \Gs{S}$ along a generalised vector $V\in \Gs{E}$ by
\begin{equation}
\label{eq:Dorfman-Hd}
	\LgenS_V \epsilon = \Dgen_V \epsilon - (\Dgen \proj{\adj{P}} V) \cdot \epsilon ,
\end{equation}
where $\Dgen$ is any torsion-free $\dHd$ compatible generalised connection and $P$ is the $\dHd$ principal bundle. 

For concretness, let us work out explicitly in indices the form of the KD derivative in a few different generalised geometries. For the original formulation, based on a $TM\oplus T^*M$ generalised tangent space with a local structure group $\Spin(d)\times\Spin(d)$, we use the notation of~\cite{CSW1} and find that for a generalised vector $V^A =(V^a, \,V^{\bar{a}})$ and a spinor $\epsilon = (\epsilon^+,\,\epsilon^-)$ the derivative reads
\begin{equation}
\begin{aligned}
	 \LgenS_V \epsilon^+
		&=  V^{a} \Dgen_{a}\epsilon^+ + V^{\bar{a}} \Dgen_{\bar{a}}\epsilon^+
		+ \tfrac12 ( \Dgen_a V_b )\gamma^{ab}\epsilon^+ ,\\
	 \LgenS_V \epsilon^-
		&=  V^{a} \Dgen_{a}\epsilon^- + V^{\bar{a}} \Dgen_{\bar{a}}\epsilon^-
		+ \tfrac12 ( \Dgen_{\bar{a}} V_{\bar{b}} )\gamma^{\bar{a}\bar{b}}\epsilon^-.
		\end{aligned}
\end{equation}

For $E_{7(7)}\times\bbR^+$ generalised geometry, which is the case we will be mostly focused on in the following sections, we have an $\tilde{H}_7 = \SU(8)$ expression, and for a generalised vector $V=(V^{\alpha\alpha'}, \bar{V}_{\alpha\alpha'})$ we find,
\begin{equation}
\label{eq:gen-KD-su8}
	\LgenS_V \epsilon^\alpha 
		= \tfrac{1}{32} (V^{\gamma \gamma'} \bar\Dgen_{\gamma\gamma'} 
			+ \bar{V}_{\gamma \gamma'} \Dgen^{\gamma\gamma'}) \epsilon^\alpha
			- \tfrac{1}{16}\Big[ \Big( \bar\Dgen_{\beta \gamma} V^{\alpha \gamma}
				-  \Dgen^{\alpha \gamma} \bar{V}_{\beta \gamma} \Big)
			-\tfrac18 \delta^\alpha{}_\beta 
				\Big(\bar\Dgen_{\gamma \gamma'} V^{\gamma \gamma'}
				-  \Dgen^{\gamma \gamma'} \bar{V}_{\gamma \gamma'} \Big) \Big]
			\epsilon^\beta .
\end{equation}
Here, and throughout this paper, we are using the $\SU(8)$ index conventions of~\cite{CSW3, LSW2}.

Similar expressions can be found for the lower rank exceptional geometries. For $E_{6(6)}\times\bbR^+$, $\tilde{H}_6 = \Symp(8)$ and the generalised vector $V = (V^{[\alpha\beta]})$ transforming in the $\rep{27}$ representation we have
\begin{equation}
\label{eq:gen-KD-sp8}
	\LgenS_V \epsilon^\alpha 
		= \tfrac12 V^{\gamma \gamma'} \Dgen_{\gamma\gamma'} \epsilon^\alpha
			+ (\Dgen^{\alpha\gamma} V_{\beta\gamma} 
				+ \Dgen_\beta{}^\gamma V^\alpha{}_\gamma) \epsilon^\beta .
\end{equation}
Similarly, the generalised vector of $E_{5(5)}\times\bbR^+$ generalised geometry transforms in the $\repp{4}{4}$ representation of the compact subgroup $\tilde{H}_5 = \Spin(5) \times \Spin(5)$ and we have
\begin{equation}
\begin{aligned}
	 \LgenS_V \epsilon^{+\alpha}
		&=  V^{\beta \bar\gamma} \Dgen_{\beta\bar\gamma} \epsilon^{+\alpha} 
			-\tfrac12 (\Dgen^\alpha{}_{\bar\gamma} V_\beta{}^{\bar\gamma}
				+ \Dgen_{\beta\bar\gamma} V^{\alpha\bar\gamma})
			\epsilon^{+\beta} ,\\
	 \LgenS_V \epsilon^{-\bar\alpha}
		&=  V^{\beta \bar\gamma} \Dgen_{\beta\bar\gamma} \epsilon^{-\bar\alpha} 
		-\tfrac12 (\Dgen_\gamma{}^{\bar\alpha} V^\gamma{}_{\bar\beta}
				+ \Dgen_{\gamma\bar\beta} V^{\gamma\bar\alpha})
		\epsilon^{-\bar\beta} .
		\end{aligned}
\end{equation}
Finally, in $E_{4(4)}\times\bbR^+$ generalised geometry~\cite{CSW3}, with $\tilde{H}_4 = \Spin(5)$ and along a generalised vector $V^{ab}=V^{[ab]}$, the bracket reads
\begin{equation}
\label{eq:SO5-spinor-Dorfman}
	\LgenS_{V} \epsilon = V^{ab}\Dgen_{ab} \epsilon + \tfrac12 (\Dgen_{ac} V_b{}^c) \gamma^{ab} \epsilon.
\end{equation}

This derivative clearly has a natural action on arbitrary generalised $\dHd$ tensors as well. However, analogously to the usual Kosmann derivative, the closure of this bracket only holds for generalised Killing vectors, i.e.
\begin{equation}
\label{eq:Kosmann-closure}
	\left[\LgenS_V, \LgenS_{W}\right]\epsilon = \LgenS_{\LgenS_V W} \epsilon  \iff \Lgen_V G = \Lgen_{W} G = 0 .
\end{equation}

Note that, as mentioned before, a generalised Killing vector field $V$ satisfies~\eqref{eq:GKV-ad-proj}, and therefore the Dorfman derivative (which acts on $\Edd\times\bbR^+$ tensors, which we now think of as $\dHd$ tensors) along a GKV coincides with the KD derivative
\begin{equation}
\label{eq:Dorfman-GKV}
	\Lgen_V  = \Dgen_V   - (\Dgen \proj{\adj{\tilde{F}}} V) \cdot  = \Dgen_V - (\Dgen \proj{\adj{P}} V) \cdot = \LgenS_V .
\end{equation}
We also find, similarly to~\eqref{eq:Dorfman-GKV-Lv}, that on a non-conformal split frame a GKV satisfies the relation
\begin{equation}
\label{eq:spinor-Dorfman-GKV}
	\LgenS_V \epsilon = \mathcal{L}^{{}^\text{K}}_v \epsilon ,
\end{equation}
where $\mathcal{L}^{{}^\text{K}}_v \epsilon$ is the ordinary spinor Kosmann-Lie derivative. We provide some details on the derivation of this extremely useful lemma in appendix~\ref{app:spinor-Dorfman-GKV}.

\section{The Killing superalgebra in generalised geometry}
\label{sec:KSA}

We will now derive the result that the Killing spinors of the internal space of generic Minkowski backgrounds generate an algebra, using the tools of generalised geometry. As usual, we work in $d=7$ but all the computations are straightforward to reproduce in other dimensions, see~\cite{CSW2,CSW3}.


\subsection{An internal sector of the algebra in generalised geometry}
\label{sec:gen-kill-al}

In $E_{7(7)}\times\bbR^+$ generalised geometry, the rescaled seven-dimensional complex spinor fields $\epsilon$ defined in~\eqref{eq:rescaled-spinor} are promoted to sections of a generalised spin bundle $S$ transforming in the $\rep{8}$ representation of the enlarged local symmetry group $\SU(8)$. Using a torsion-free $\SU(8)$ compatible generalised connection $\Dgen$, it was shown in~\cite{CSW3,CSW4} that the Killing spinor equations~\eqref{eq:7d-susy} could be rewritten in generalised geometry language as the projections of the generalised derivative of the spinor
\begin{equation}
\label{eq:susy-su8}
\begin{aligned}
	(\Dgen \proj{J} \epsilon){}^{[\alpha\beta\gamma]} &= \Dgen^{[\alpha\beta}\epsilon^{\gamma]} = 0,\\
	(\Dgen \proj{{S}} \epsilon){}_\alpha 
		&= -\Dgen_{\alpha\beta} \epsilon^\beta = 0 .
\end{aligned}
\end{equation}
where $J$ in $d=7$ corresponds to the $\rep{56}$ representation, see table~\ref{table}.

We can then, for instance, re-write  the relation~\eqref{eq:comm-susy-lie} for a spinor $\zeta\in\Gs{S}$.
Recall that this projected derivative $ \Dgen \proj{S\oplus J}\zeta$ for a torsion-free metric-compatible connection is such that it is uniquely determined by the supergravity fields, or to use a more generalised geometry terminology, it depends only on the generalised metric $G$. As such, the action of the KD derivative along a GKV will necessarily commute with it, i.e. given a $V\in\Gs{E}$ such that $\LgenS_V G = 0$, we have the $\dHd$-covariant formula
\begin{equation}
\label{eq:gen-comm-susy-lie}
	\LgenS_{V} \left( \Dgen \proj{S\oplus J}\zeta\right) = \Dgen \proj{S\oplus J} \left(\LgenS_{V} \zeta\right).
\end{equation}
One can double-check this by going to a non-conformal split frame, in which, thanks to the lemma~\eqref{eq:spinor-Dorfman-GKV}, the KD bracket reduces to the Lie bracket and we recover precisely~\eqref{eq:comm-susy-lie}.

Considering the basis of orthonormal Killing spinors $\epsilon_i$ with $i = 1,\dots , \mathcal{N}$ introduced in~\eqref{eq:orth-sp}, we can easily construct a set of complex generalised vectors $V_{ij}$ and $W^{ij}$ as 
\begin{equation}
\label{eq:gen-bilinears}
\begin{aligned}
	(V_{ij})^{\alpha \beta} &= \epsilon^{[\alpha}_i \epsilon^{\beta]}_j ,
	 & \hs{40pt} 
	(V_{ij})_{\alpha \beta} &= 0 ,\\
	(W^{ij})^{\alpha \beta} &= 0 ,
	 & \hs{40pt} 
	(W^{ij})_{\alpha \beta} &= \bar{\epsilon}^i_{[\alpha} \bar{\epsilon}^j_{\beta]} .
\end{aligned}
\end{equation}
Note that the $\SU(8)$ indices are defined with respect to the conformal split frame, so that when we examine the vector part $v_{ij}$ of $V_{ij}$ in the coordinate basis, we find
\begin{equation}
	(v_{ij})^m \sim \hat\epsilon^c_i \gamma^m \hat\epsilon_j ,
\end{equation}
which would be the natural spinor bilinear from the supergravity perspective.

It is easy to see that the $V_{ij}$ and $W^{ij}$ are generalised Killing vectors. From~\eqref{eq:susy-su8}, we can immediately observe that the components of $\Dgen \proj{\adj{P}^\perp} V_{ij}$ vanish
\begin{equation}
\label{eq:ks-gkv}
	\Dgen^{[\alpha \beta} (\epsilon_i^\gamma \epsilon_j^{\delta]}) = 0 ,
	\hs{50pt}
	\Dgen_{\alpha \beta}(\epsilon_i^{[\alpha} \epsilon_j^{\beta]}) = 0 ,
\end{equation}
and similarly for $W^{ij}$. Thus from~\eqref{eq:GKV-ad-proj}, $V_{ij}$ and $W^{ij}$ are GKVs. Now, from~\eqref{eq:gen-comm-susy-lie} we have that the KD derivative of a Killing spinor by a GKV is itself a Killing spinor, so we can introduce constant coefficients $X_{ijk}{}^l$ such that
\begin{equation}
	\LgenS_{V_{ij}} \epsilon_k = X_{ijk}{}^l \epsilon_l , 
\end{equation}
and similarly for $W^{ij}$.

We will now use the explicit form of the KD derivative to show that supersymmetry implies that the constant coefficients $X_{ijk}{}^l$ must vanish. As we saw earlier, the expression for the KD derivative~\eqref{eq:Dorfman-Hd} of a spinor by a generalised vector reads
\begin{equation}
\label{eq:spinor-Dorfman-su8}
	 \LgenS_V \epsilon^\alpha 
		= \tfrac{1}{32}(V^{\gamma \gamma'} \bar\Dgen_{\gamma\gamma'} 
			+ \bar{V}_{\gamma \gamma'} \Dgen^{\gamma\gamma'}) \epsilon^\alpha
			- \tfrac{1}{16}\Big[ \Big( \bar\Dgen_{\beta \gamma} V^{\alpha \gamma}
				-  \Dgen^{\alpha \gamma} \bar{V}_{\beta \gamma} \Big)
			-\tfrac18 \delta^\alpha{}_\beta 
				\Big(\bar\Dgen_{\gamma \gamma'} V^{\gamma \gamma'}
				-  \Dgen^{\gamma \gamma'} \bar{V}_{\gamma \gamma'} \Big) \Big]
			\epsilon^\beta .
\end{equation}
Substituting in $(V_{ij})^{\alpha \beta} = \epsilon^{[\alpha}_i \epsilon^{\beta]}_j$, $
	(\bar{V}_{ij})_{\alpha \beta} = 0$ we find
\begin{equation*}
\begin{aligned}
	 \LgenS_{V_{ij}} \epsilon_k^\alpha 
		&= \tfrac{1}{32} \epsilon_{i}^{\gamma} \epsilon_{j}^{\gamma'} \bar\Dgen_{\gamma\gamma'} \epsilon_k^\alpha
			- \tfrac{1}{16}\epsilon_k^\beta \left(\bar\Dgen_{\beta \gamma} \epsilon_{i}^{[\alpha} \epsilon_{j}^{\gamma]}\right) - \tfrac{1}{128} \epsilon_k^\alpha \left(\bar\Dgen_{\gamma\gamma'}\epsilon_{i}^{\gamma} \epsilon_{j}^{\gamma'}\right)  \\
			 &= \tfrac{1}{32} \epsilon_{i}^{\gamma} \epsilon_{j}^{\gamma'} \bar\Dgen_{\gamma\gamma'} \epsilon_k^\alpha
			- \tfrac{1}{32} \epsilon_k^\beta \epsilon_{j}^{\gamma} \bar\Dgen_{\beta \gamma} \epsilon_{i}^{\alpha} 
			- \tfrac{1}{32} \epsilon_k^\beta \epsilon_{i}^{\gamma}  \bar\Dgen_{\beta \gamma} \epsilon_{j}^{\alpha},
\end{aligned}
\end{equation*}
where in the last equality we used repeatedly the Killing spinor equation $\Dgen_{\alpha\beta} \epsilon^\beta = 0 $. On the other hand, for $(W^{ij})^{\alpha \beta} = 0$, $(\bar W^{ij})_{\alpha \beta} = \bar{\epsilon}^i_{[\alpha} \bar{\epsilon}^j_{\beta]}$ we obtain 
\begin{equation*}
\begin{aligned}
	 \LgenS_{W_{ij}} \epsilon_k^\alpha 
		&= \tfrac{1}{32} \bar{\epsilon}^i_{\gamma} \bar{\epsilon}^j_{\gamma'} \Dgen^{\gamma\gamma'} \epsilon_k^\alpha
			+\tfrac{1}{16} \epsilon_k^\beta \left( \Dgen^{\alpha \gamma} \bar{\epsilon}^i_{[\beta} \bar{\epsilon}^j_{\gamma]}\right) 
			+\tfrac{1}{128} \epsilon_k^\alpha\left( \Dgen^{\gamma\gamma'}\bar{\epsilon}^i_{\gamma} \bar{\epsilon}^j_{\gamma'}\right) \\
			 &= -\tfrac{1}{32} \bar{\epsilon}^i_{\gamma} \bar{\epsilon}^j_{\gamma'} \Dgen^{\alpha\gamma} \epsilon_k^{\gamma'}  -\tfrac{1}{32} \bar{\epsilon}^i_{\gamma} \bar{\epsilon}^j_{\gamma'} \Dgen^{\gamma'\alpha} \epsilon_k^\gamma
			+ \tfrac{1}{32}\epsilon_k^\beta \bar{\epsilon}^j_{\gamma} \Dgen^{\alpha \gamma} \bar{\epsilon}^i_{\beta}  - \tfrac{1}{32}\epsilon_k^\beta \bar{\epsilon}^i_{\gamma} \Dgen^{\alpha \gamma} \bar{\epsilon}^j_{\beta}  		\\
			&=	-\tfrac{1}{32} \bar{\epsilon}^i_{\gamma} \left(\Dgen^{\alpha\gamma}\bar{\epsilon}^j_{\gamma'}  \epsilon_k^{\gamma'} \right) -\tfrac{1}{32}  \bar{\epsilon}^j_{\gamma'} \left(\Dgen^{\gamma'\alpha} \bar{\epsilon}^i_{\gamma}\epsilon_k^\gamma\right) = 0,
\end{aligned}
\end{equation*}
where we first used both of the supersymmetry conditions~\eqref{eq:susy-su8} and then finally the orthonormality of the basis $\epsilon_i$. We therefore conclude
\begin{equation}
\label{eq:trilinear-operator}
	 (\LgenS_{V_{ij}} \epsilon_k)^\alpha
		= \tfrac{3}{32} \epsilon_{[i}^{\gamma} \epsilon_{j}^{\gamma'}
			( \bar\Dgen_{|\gamma \gamma'|} \epsilon_{k]}^{\alpha}) ,
	\hs{40pt}
	\LgenS_{W^{ij}} \epsilon_k = 0 .
\end{equation}
Note that the RHS of the first of these equations would be automatically vanishing for $\mathcal{N} \leq 2$ by anti-symmetry. However, again using that we have an orthonormal basis, we can write
\begin{equation}
	X_{ijk}{}^l = \bar{\epsilon}{}^l{}_\alpha (\LgenS_{V_{ij}} \epsilon_k){}^\alpha
		= - \tfrac{3}{32} \epsilon_{[i}^{\gamma} \epsilon_{j}^{\gamma'} 
		\epsilon_{k]}^{\alpha}
			( \bar\Dgen_{\gamma \gamma'} \bar{\epsilon}{}^l{}_\alpha)
		= - \tfrac{3}{32} \epsilon_{[i}^{\alpha} \epsilon_{j}^{\beta} 
		\epsilon_{k]}^{\gamma}
			( \bar\Dgen_{[\alpha \beta} \bar{\epsilon}{}^l{}_{\gamma]}) = 0  ,
\end{equation}
using the complex conjugate of the first supersymmetry condition in~\eqref{eq:susy-su8}. We have now arrived at a key result: the coefficients $X_{ijk}{}^l$ vanish and we have
\begin{equation}
\label{eq:vanishing-trilinears}
	\LgenS_{V_{ij}} \epsilon_k = 0 ,
	\hs{40pt}
	\LgenS_{W^{ij}} \epsilon_k = 0 .
\end{equation}
%


\subsection{The Killing superalgebra}

Putting all of this together, we arrive at an algebra very much like the eleven-dimensional Killing superalgebra, but in the internal sector of the dimensional split.
If we define $\mathfrak{g}_0$ as the space of generalised Killing vectors and $\mathfrak{g}_1$ the space of Killing spinors, we have that $\mathfrak{g}_0\oplus\mathfrak{g}_1$ equipped with the Kosmann-Dorfman bracket forms an algebra. We have that the Dorfman bracket on $\mathfrak{g}_0 \times \mathfrak{g}_0 \rightarrow \mathfrak{g}_0$ gives an algebra by itself since it satisfies Jacobi and is Leibniz, so in particular the Dorfman derivative of a GKV is itself a GKV. The map $\mathfrak{g}_1\times\mathfrak{g}_1\rightarrow\mathfrak{g}_0$ is simply~\eqref{eq:gen-bilinears} thanks to~\eqref{eq:ks-gkv}. The map $\mathfrak{g}_0\times\mathfrak{g}_1\rightarrow\mathfrak{g}_1$ is the Kosmann-Dorfman bracket since it satisfies~\eqref{eq:gen-comm-susy-lie} and the Jacobi identity for $[\mathfrak{g}_0,\mathfrak{g}_0,\mathfrak{g}_1]$ is ensured by~\eqref{eq:Kosmann-closure}. Finally, we have just computed~\eqref{eq:vanishing-trilinears}, which means that the remaining Jacobi identity  [$\mathfrak{g}_1,\mathfrak{g}_1,\mathfrak{g}_1]$ is satisfied trivially.

Looking just at the ideal $[\mathfrak{g}_1,\mathfrak{g}_1]\oplus\mathfrak{g}_1$, we can also calculate the Dorfman derivative algebra of the generalised vectors $V_{ij}$ and $W^{ij}$. Using the Leibniz property of the Dorfman derivative we find
\begin{equation}
\begin{aligned}
	\Lgen_{V_{ij}} V_{kl} &= 0 ,
	& \hs{40pt}
	\Lgen_{W^{ij}} V_{kl} &= 0 ,\\
	\Lgen_{V_{ij}} W^{kl} &= 0 ,
	& \hs{40pt}
	\Lgen_{W^{ij}} W^{kl} &= 0 ,\\
\end{aligned}
\end{equation}
so we have the particularly simple subalgebra generated by just the Killing spinors
\begin{equation}
\label{eq:gen-KSA}
\begin{aligned}[]
	[ \epsilon_i, \epsilon_j ] &= V_{ij} ,
	& \hs{40pt} 
	[ \bar{\epsilon}{}^i, \bar{\epsilon}{}^j ] &= W^{ij} ,\\
	[ V_{ij} , \epsilon_k ] &= \LgenS_{V_{ij}} \epsilon_k = 0 ,
	& \hs{40pt}
	[ W^{ij} , \epsilon_k ] &= \LgenS_{W^{ij}} \epsilon_k = 0 ,\\
	[ V_{ij} , \bar{\epsilon}{}^k ] &= \LgenS_{V_{ij}} \bar{\epsilon}{}^k = 0 ,
	& \hs{40pt} 
	[ W^{ij} , \bar{\epsilon}{}^k ] &= \LgenS_{W^{ij}} \bar{\epsilon}{}^k = 0 ,\\
	[V_{ii'}, V_{jj'} ]  &= \Lgen_{V_{ii'}} V_{jj'} = 0 ,
	& \hs{40pt} 
	[W^{ii'}, W^{jj'} ]  &= \Lgen_{W^{ii'}} W^{jj'} = 0 ,\\
	& & [V_{ii'}, W^{jj'} ]  = \Lgen_{V_{ii'}} W^{jj'} = 0 .& 
\end{aligned}
\end{equation}
Note that all the brackets here are naturally anti-symmetric so that we have found a Lie algebra rather than a Lie superalgebra.


\subsection{Eleven-dimensional interpretation}

Let us now briefly consider the implications of~\eqref{eq:gen-KSA} for the eleven-dimensional Killing superalgebra of such backgrounds. Equation~\eqref{eq:spinor-decomp} should strictly be viewed as giving $\mathcal{N}$ different embeddings of the constant four-dimensional Majorana spinors $\eta$ (i.e. the Killing spinors of Minkowski space) into the eleven-dimensional Killing spinors. Specifically, each internal Killing spinor $\hat\epsilon_i$ in the basis provides us with a linear map
\begin{equation}
\begin{aligned}
\label{eq:spinor-embed}
	\varepsilon_i :	\eta &\longmapsto \varepsilon_i(\eta) 
		= \eta^+ \otimes \hat\epsilon_i + \eta^- \otimes (\tilde{D} \hat\epsilon_i)^* .
\end{aligned}
\end{equation}
Then the vector $v_{ij, \eta_1 \eta_2} = v\big(\varepsilon_i(\eta_1), \varepsilon_j(\eta_2)\big)$ which is defined by the eleven-dimensional spinors will satisfy, by~\eqref{eq:vanishing-trilinears},
\begin{equation}
	\mathcal{L}_{v_{ij, \eta_1 \eta_2}} \varepsilon_{k,\eta_3} = 0 .
\end{equation}
We also have that $v_{ij, \eta_1 \eta_2} $ is purely internal if $i\neq j$, with components
\begin{equation}
	(v_{ij, \eta_1 \eta_2})^m 
		=  (\eta_{1+}^T \tilde{C} \eta_{2+}) (\hat\epsilon^c_i \gamma^m \hat\epsilon_j) + (\text{c.c.}) ,
\end{equation}
while for $i=j$ we have a purely external vector with\footnote{Note that $\Gamma^\mu$ with an eleven-dimensional coordinate index includes a factor of $\ee^{-\Delta}$.}
\begin{equation}
	(v_{ii, \eta_1 \eta_2})^\mu 
		=  \ee^{-\Delta}(\eta_{1+}^T \tilde{C} \gamma^\mu \eta_{2-}) (\hat\epsilon^\dagger_i \hat\epsilon_i)^* + (\text{c.c.})
		= (\eta_{1}^T \tilde{C} \gamma^\mu \eta_{2}) .
\end{equation}
Adopting a notation $p_{\eta_1 \eta_2} = v_{ii, \eta_1 \eta_2}$ for $i=j$, while setting $z_{ij,\eta_1 \eta_2} = v_{ij, \eta_1 \eta_2}$ for $i\neq j$, we have that the Killing superalgebra generated by the Killing spinors $\varepsilon_{i,\eta}$ is then given by
\begin{equation}
\label{eq:gen-KSA-11d}
\begin{aligned}[]
	[ \varepsilon_{i,\eta_1}, \varepsilon_{j,\eta_2} \} &= v_{ij, \eta_1 \eta_2} 
		= \delta_{ij} p_{\eta_1 \eta_2} + z_{ij,\eta_1 \eta_2}, \\
	[ v_{ij, \eta_1 \eta_2} , \varepsilon_{k,\eta_3} \} &=  0 , \\
	[ v_{ij, \eta_1 \eta_2} , v_{kl, \eta_3 \eta_4} \} &=  0 .
\end{aligned}
\end{equation}
This is the supertranslational part of the $\mathcal{N}$-extended super-Poincar\'e algebra with central charges given by the internal vectors $z_{ij,\eta_1 \eta_2}$, which generate isometries of the internal background. 

One can view these formulae as giving an uplift of the $\mathcal{N}$-extended super-Poincar\'e algebra on the external Minkowski space into the eleven-dimensional theory. Working in the Weyl basis for the four-dimensional external space gamma matrices, we can introduce a canonical basis of Weyl spinors $\eta_1 = (1,0)$, $\eta_2 = (0,1)$ (along with the conjugates $\bar\eta^{\dot{1}} = (1,0)$, $\bar\eta^{\dot{2}} = (0,1)$) to define
\begin{equation}
\begin{aligned}[]
	Q_{i,\alpha} = \eta_\alpha \otimes \hat\epsilon_i ,
	\hs{40pt}
	\bar{Q}_i{}^{\dot\alpha}  = \bar{\eta}^{\dot\alpha} \otimes \hat\epsilon^c_i ,
\end{aligned}
\end{equation}
and write, for example,
\begin{equation}
\begin{aligned}[]
	z_{ij,\eta_1 \eta_2} = z_{ij} \epsilon_{\alpha\beta} \eta^{\alpha}_{1+} \eta^\beta_{2+}
		+ \bar{z}_{ij} \epsilon_{\dot\alpha \dot\beta} \eta^{\dot\alpha}_{1-} \eta^{\dot\beta}_{2-} ,
\end{aligned}
\end{equation}
where $z^m_{ij} = \hat\epsilon_i^c \gamma^m \hat\epsilon_j$ and we are using the usual $\SL(2,\bbC)$ Weyl-spinor index notation. We then have also $\varepsilon_{i,\eta} = Q_{i,\alpha} \eta_+^\alpha + \bar{Q}_i{}^{\dot\alpha}\eta^-_{\dot\alpha}$ and find that the first line of~\eqref{eq:gen-KSA-11d} takes the familiar form
\begin{equation}
\label{eq:gen-KSA-11d-Weyl}
\begin{aligned}[]
	[ Q_{i,\alpha}, \bar{Q}_{j,\dot\beta} \}
		&= \delta_{ij} (\sigma^\mu)_{\alpha \dot\beta} \tfrac{\der}{\der x^\mu} ,\\
	[ Q_{i,\alpha}, Q_{j,\beta} \}
		&=  \epsilon_{\alpha\beta} z_{ij} , \\
	[ \bar{Q}_{i,\dot\alpha}, \bar{Q}_{j,\dot\beta} \}
		&=  \epsilon_{\dot\alpha \dot\beta} \bar{z}_{ij} ,
\end{aligned}
\end{equation}
and we also have
\begin{equation}
\label{eq:gen-KSA-11d-Weyl2}
\begin{aligned}[]
	[ z_{ij}, z_{kl} \} = [ z_{ij}, Q_{k,\alpha} \}
		&= [ z_{ij} \bar{Q}_{k,\dot\alpha} \} = 0 ,\\
	[ \bar{z}_{ij},  \bar{z}_{kl} \} = [ \bar{z}_{ij}, Q_{k,\alpha} \}
		&=  [ \bar{z}_{ij}, \bar{Q}_{k,\dot\alpha} \} = 0 . \\
\end{aligned}
\end{equation}
%


\section{Generalised holonomy for supersymmetric backgrounds}
\label{sec:gen-hol}

In~\cite{CSW4} two crucial results were established. First, that the existence of $\mathcal{N}$ independent spinor fields on the internal space $M$ defines a reduction of the structure group on the generalised tangent bundle to the groups $G_{\mathcal{N}}$ listed on table~\ref{table2}. Secondly, and more remarkably, it was proven that for $\mathcal{N}=1$ the generalised intrinsic torsion of these structures vanishes if and only if the spinor satisfies the Killing spinor equations. Such backgrounds were dubbed generalised special holonomy spaces, in analogy to the usual
special holonomy manifolds which arise in fluxless supersymmetric compactifications.

\begin{table}[htb]
\centering
\begin{tabular}{lll}
   $d$ & $\dHd$ &  $\gN$  \\ 
   \hline
   7 & $\SU(8)$ &  $\SU(8-\mathcal{N})$\\
   6 & $ \Symp(8)$ & $\Symp(8-2\mathcal{N})$ \\
   5 & $\Symp(4)\!\! \times \!\! \Symp(4)$ &   
      $\Symp(4-2\mathcal{N}_+)\!\times\!\Symp(4-2\mathcal{N}_-)$\\
   4 & $\Symp(4)$ & $\Symp(4-2\mathcal{N}) $
\end{tabular}
\caption{Generalised structure subgroups $G_{\mathcal{N}}\subset \dHd$ preserving $\mathcal{N}$ supersymmetry in $(11-d)$-dimensional Minkowski backgrounds. Note that for $d = 5$ we have six-dimensional supergravity with $(\mathcal{N}_+,\mathcal{N}_-)$ supersymmetry.}
\label{table2}
\end{table}

In the following, we will show that precisely the same statement can be made for backgrounds with more supersymmetry, that is we can conclude that
\begin{quote}
   \textit{The internal spaces of supersymmetric Minkowski backgrounds are precisely the spaces of generalised $\gN$ special holonomy.} 
\end{quote}
We remark that (for Euclidean signature) these spaces are automatically generalised Ricci-flat, or in supergravity language, they solve the equations of motion, by the argument of~\cite{CSW4}.

The methods we will use can be reproduced in any dimension $d < 8$, though for concreteness we will work only on the case of a four-dimensional external Minkowski space, that is with $d=7$ and for which the Killing spinors $\epsilon_i$ define a global $\SU(8-\mathcal{N})$ structure on the generalised tangent space. The task is to show that this $\SU(8-\mathcal{N})$ structure has vanishing intrinsic torsion, or equivalently that there exists a torsion-free generalised spin connection $\hat\Dgen$ with respect to which the $\epsilon_i$ are parallel
\begin{equation}
	\hat\Dgen \epsilon_i = 0 .
\end{equation}

In the following, for the sake of readability we will use a slight abuse of notation in which we identify bundles with their corresponding representations. 


\subsection{Generalised intrinsic torsion for $\gN$ structures}
\label{sec:gen-sp-hol}

Let $\Dgen$ and $\hat\Dgen$ be any two generalised spin connections. In $d=7$ these are $\SU(8)$ compatible connections. The difference between the two defines a tensor $\hat\Sigma=\hat\Dgen-\Dgen$ taking values in the bundle $\HConSp := E^* \otimes \adj P_{\SU(8)}$. In spinor indices, these are given by
\begin{equation}
   \hat\Sigma = (\hat\Sigma_{\alpha\beta}{}^\gamma{}_\delta , 
        \bar{\hat\Sigma}^{\alpha\beta}{}^\gamma{}_\delta  )
        \in (\rep{28}+\bar{\rep{28}})\times\rep{63} = \Kgen_{SU(8)}, 
\end{equation}
where the elements are antisymmetric on $\alpha$ and $\beta$ and traceless on contracting $\gamma$ with $\delta$. 

Denoting the generalised torsion of $\Dgen$ by $T(D)$, we can define a map $\tau$ between the space of connections $\HConSp$ and the space of torsions $W$
\begin{equation}
\label{eq:tau}
\begin{aligned}
	\tau : \HConSp &\ra W ,\\
		\hat\Sigma &\mapsto T(\hat{D}) - T({D}) .
\end{aligned}
\end{equation}
Up to overall normalisations, this map is given by~\cite{CSW3,CSW4} 
\begin{equation}
\begin{aligned}
\label{eq:torsion-su8}
   \tau(\hat\Sigma)_{\alpha\beta} 
      &= \hat\Sigma_{\alpha \gamma }{}^\gamma{}_{\beta } , 
      && \in \rep{28} + \rep{36} , \\
   \tau(\hat\Sigma)_{\alpha\beta\gamma}{}^\delta 
      &= \hat\Sigma^0_{[\alpha\beta}{}^\delta{}_{\gamma]} = \hat\Sigma_{[\alpha\beta}{}^\delta{}_{\gamma]} + \tfrac13 \hat\Sigma_{[\alpha |\epsilon |}{}^\epsilon{}_{\beta } \delta^\delta{}_{\gamma]} , 
      && \in \rep{420} , 
\end{aligned}
\end{equation}
where the ``0'' superscript on $\hat\Sigma^0_{[\alpha\beta}{}^\delta{}_{\gamma]}$ means it is completely traceless and there are similar expressions for the conjugate representations in terms of $\bar{\hat\Sigma}$. 

We now assume we have $\mathcal{N}$ independent spinors defining a $\gN = \SU(8-\mathcal{N})$ structure $P_{\gN}$ on the generalised tangent bundle. We can then decompose the $\SU(8)$ torsion representations under $\gN$, and these are listed in appendix~\ref{app:torsion}.

A tensor $\Sigma$ defined by connections which we require to be compatible with $P_{\gN}$ will be an element of a restricted subspace $\Kgen_{\gN} := E^* \otimes \adj P_{\gN}\subset \HConSp$. If we split the spinor indices $\alpha$ into $a=1,\dots,8-\mathcal{N}$ and $i=1,\dots,\mathcal{N}$ we find that its non-zero components are
\begin{equation}
\begin{aligned}
  &\Sigma_{ab}{}^c{}_d  , \\
   &\Sigma_{ai}{}^c{}_d = -\Sigma_{ia}{}^c{}_d
        , \\
   &\Sigma_{ij}{}^c{}_d  ,    
\end{aligned}
\end{equation}
and similarly for the conjugate $\bar{\Sigma}$. The corresponding $\gN$ representations are listed in appendix~\ref{app:comp-c}. 

Now, a quick examination of the representations listed in the appendices reveals that there are those that appear in the decomposition of the torsion space $W$ but not in that of $\Kgen_{\gN}$. This means that the image of the restricted map $\tau|_{\Kgen_{\gN}}:=\tau_{\gN}$ does not fill out the entire torsion space. Denoting this image  $\text{Im}\,\tau_{\gN} := W_{\gN} $, we can therefore define another bundle, the space of generalised intrinsic torsions
\begin{equation}
	W_{\text{int}} = \frac{W}{W_{\gN}}.
\end{equation}
The representations which make up $\Tint$ are listed in appendix~\ref{app:int-t}.
The torsion of any $\gN$ compatible $D$ naturally projects onto $\Tint$ -- this is the generalised intrinsic torsion $T_{\text{int}} := T(D)|_{W_{\text{int}}}$ which, by construction, is independent of $\Sigma$ and thus is common to all $\gN$ connections. A non-zero $T_{\text{int}}$ measures the obstruction to finding a torsion-free generalised connection which preserves the $\gN$ structure.

We can find this projection explicitly in indices (see appendix~\ref{app:int-torsion-calc} for a more detailed derivation). Applying the $\tau$ map to $\Sigma$, we obtain that the following (reducible) components are unconstrained
\begin{equation}
\label{eq:map2}
\begin{aligned}
   \tau(\Sigma)_{ab} &= \Sigma_{ac}{}^c{}_{b}   , & 
      && && &&  
   \tau(\Sigma)_{ia}  &= \Sigma_{ic}{}^c{}_{a}  , \\
      \tau(\Sigma)_{abc}{}^d &= \Sigma_{[ab}{}^d{}_{c]}  + \tfrac13 \Sigma_{[a|e|}{}^e{}_{b} \delta^d{}_{c]} ,
        & 
      && && && 
  \tau(\Sigma)_{abi}{}^c &= \tfrac23\Sigma_{i[a}{}^c{}_{b]} +\tfrac29  \Sigma_{ie}{}^e{}_{[a} \delta^c{}_{b]} 
      ,\\
     \tau(\Sigma)_{ija}{}^b & = \tfrac13 \Sigma_{ij}{}^b{}_{a}   
     ,  
\end{aligned}
\end{equation}
which therefore give the image of $\tau_{\gN}$, while the remaining combinations of components are fixed whatever the choice of compatible $\Sigma$
\begin{equation}
\label{eq:int-torsion-red}
\begin{aligned}
      \tau(\Sigma)_{ij}  &=0, & 
      && && &&  &\tau(\Sigma)_{ai} = 0 
     , \\
     \tau(\Sigma)_{abc}{}^i &= 0   , & 
      && && && &\tau(\Sigma)_{abi}{}^j = \tfrac19 \Sigma_{[a|e|}{}^e{}_{b]} \delta^i{}_{j} = \tfrac19\tau(\Sigma)_{[ab]}\delta^i{}_{j}
     , \\
    \tau(\Sigma)_{ijk}{}^l &= 0 , & 
      && && &&  &\tau(\Sigma)_{aij}{}^k = -\tfrac19  \Sigma_{[i|e}{}^e{}_{a|} \delta^k{}_{j]}  = -\tfrac19\tau(\Sigma)_{[i|a|} \delta^k{}_{j]} ,
    \\
   \tau(\Sigma)_{ijk}{}^a &= 0 , &  
 && && &&
\end{aligned}
\end{equation}
and thus give, together with their conjugates, the projections to the intrinsic torsion representations -- no matter our choice of $\Sigma$, the torsion of any connection lying in those components cannot be shifted. Note however that many of these terms include anti-symmetrisations and so vanish identically for certain values of $\mathcal{N}$. A little more work is required if one wishes to rephrase these constraints in terms of genuine $SU(8-\mathcal{N})$ irreps, this is done explicitly in appendix~\ref{app:int-torsion-irrep}.
%



\subsection{Generalised special holonomy and $\mathcal{N}$ supersymmetry}
\label{sec:KSgenint}

We will now show that the supersymmetry conditions imply precisely that the intrinsic torsion of the $\gN$ structure we have just described must vanish. In $\SU(8)$ indices, we have that the set of $\mathcal{N}$ Killing spinor equations for $\mathcal{N}$ spinors $\epsilon_i$ read
\begin{equation}
\label{eq:KSE7}
   \Dgen \proj{J} \epsilon_i = \Dgen^{[\alpha\beta}\epsilon_i^{\gamma]}=0, 
   \qquad 
   \Dgen \proj{S} \epsilon_i = \bar\Dgen_{\alpha\beta}\epsilon_i^{\beta}=0, 
\end{equation}
together with their complex conjugates, and where $\Dgen$ is any torsion-free $\SU(8)$ connection. The difference between $\Dgen$ and a compatible $\gN$ connection $\hat{\Dgen}$ 
\begin{equation}
   \hat{\Dgen} = \Dgen + \hat{\Sigma} ,
\end{equation}
defines a $\hat{\Sigma}\in\Gs{\HConSp}$. Since $\Dgen$ is torsion-free we have the torsion of the $\gN$ connection
\begin{equation}
\label{eq:TS}
   T(\hat{D}) = \tau(\hat{\Sigma}) , 
\end{equation}
in terms of the $\tau$ map defined in~\eqref{eq:tau}.

Before proceeding further we must address an important subtlety --
if we were given a generic set of independent spinors $\zeta_i$ stabilised by $\gN$, we would not necessarily have that $\hat\Dgen\zeta_i = 0$. For example, while any generic non-vanishing spinor $\zeta$ defines an $\SU(7)$ structure, if its norm is non-constant we find that $\hat\Dgen \zeta = \big(\partial \log |\!|\zeta|\!|\big)\,\zeta$. Fortunately, we showed in~\eqref{eq:orth-sp} that the Killing spinors $\epsilon_i$ can be taken to be orthonormal without loss of generality, so indeed we have that $\hat{\Dgen}\epsilon_i=0$. 

In particular the projections $\hat{\Dgen}\proj{J}\epsilon_i$ and $\hat{\Dgen}\proj{S}\epsilon_i$ both vanish. Thus we have 
\begin{equation}
\label{eq:sigma-susy}
   \hat{\Sigma} \proj{J} \epsilon_i = 0 , \qquad 
   \hat{\Sigma} \proj{S} \epsilon_i = 0 .
\end{equation}
Decomposing, we have that in indices these read
\begin{equation}
\begin{aligned}
   \hat{\Sigma}_{[ab}{}^i{}_{c]} &= 0 
   , & && && 
   \hat{\Sigma}_{[ja}{}^i{}_{b]} & = 0  
   , \\ \hat{\Sigma}_{[jk}{}^i{}_{a]} & = 0  
   , & && && 
   \hat{\Sigma}_{[ij}{}^l{}_{k]} & = 0 ,\\
   \hat{\Sigma}_{ab}{}^b{}_i &= 0 
   , & && && 
   \hat{\Sigma}_{ia}{}^a{}_j  & = 0  
    , 
\end{aligned}
\end{equation}
together with their complex conjugates, with the rest of $\hat\Sigma$ left a priori unconstrained. Now we compare with the torsion map~\eqref{eq:torsion-su8} and we recognise that these projections coincide with some of its components
\begin{equation}
\begin{aligned}
   \hat{\Sigma}_{ab}{}^b{}_i &= \tau(\hat\Sigma)_{ai} = 0 , & && && \hat{\Sigma}_{ia}{}^a{}_j  & = \tau(\hat\Sigma)_{ij} = 0,
\end{aligned}
\end{equation}
from the projection to $S$, and
\begin{equation}
\begin{aligned}
  \hat{\Sigma}_{[ij}{}^l{}_{k]} &= \tau(\hat{\Sigma})_{ijk}{}^l  
  	- \tfrac13 \tau(\hat\Sigma)_{[ij} \delta^l{}_{k]}= 0 
   , & && &&  \hat{\Sigma}_{[ia}{}^j{}_{b]} &=  \tau(\hat\Sigma)_{abi}{}^j - \tfrac19\tau(\hat\Sigma)_{[ab]}\delta^i{}_{j} = 0 
   , \\ 
    \hat{\Sigma}_{[ab}{}^i{}_{c]} &= \tau(\hat{\Sigma})_{abc}{}^i = 0 , & && &&   
    \hat{\Sigma}_{[ij}{}^k{}_{a]} &= \left(\tau(\hat{\Sigma})_{aij}{}^k 
    	+ \tfrac19\tau(\hat\Sigma)_{[i|a|}\delta^k{}_{j]}\right)  \\
    	& && &&  && \qquad \qquad - \tfrac19\tau(\hat\Sigma)_{a[i}\delta^k{}_{j]}  = 0,
\end{aligned}
\end{equation}
for the $J$ projection. Altogether, these are nearly precisely the same constraints that we computed as giving the intrinsic torsion in~\eqref{eq:int-torsion-red}, with the single exception being $\tau(\hat\Sigma)_{ijk}{}^a$ (and its conjugate) which may also contribute to the intrinsic torsion but which we are missing in the Killing spinor equations. This corresponds to $\tbinom{\mathcal{N}}{3}$ copies of the  $\rep{[8-\mathcal{N}]}$ representation of $\SU(8-\mathcal{N})$. Thus the Killing spinor equations are setting nearly all of the components of the intrinsic torsion directly to zero. Note that, because of the anti-symmetrisation in the $i,j,k$ indices, this missing term vanishes identically for $\mathcal{N} < 3$ and our proof is done -- all the intrinsic torsion vanishes and we have generalised special holonomy.\footnote{Note also that the same holds for $\mathcal{N} = 8$, in this case because any term with an $a$ index vanishes identically. This proves that maximally supersymmetric backgrounds correspond to identity structures, i.e.  generalised parallelisations~\cite{LSW1}, which are torsion-free. These are necessarily flat with vanishing fluxes.}

To see that in a supersymmetric background $\tau(\hat\Sigma)_{ijk}{}^a$ still vanishes even for $\mathcal{N}\geq 3$, we recover the results from section~\ref{sec:gen-kill-al}. We found that the internal part of the Killing superalgebra can be expressed in terms of the Kosmann-Dorfman derivative, and in particular we have that
\begin{equation}
	\LgenS_{V_{ij}} \epsilon_k = 0 ,
\end{equation}
is satisfied for the GKV $(V_{ij})^{\alpha \beta} = \epsilon^{[\alpha}_i \epsilon^{\beta]}_j$, $(\bar{V}_{ij})_{\alpha \beta} = 0$ and where all spinors are Killing. Evaluating this for the torsion-free $\SU(8)$ connection
\begin{equation}
   \Dgen =  \hat{\Dgen} -  \hat{\Sigma} ,
\end{equation}
and keeping in mind that $\hat\Dgen\epsilon_i = 0$, we obtain from~\eqref{eq:trilinear-operator} that this expression is proportional precisely to our missing intrinsic torsion term
\begin{equation}
	(\LgenS_{V_{ij}} \epsilon_k)^a  = \tfrac{3}{32} \hat\Sigma_{[ij}{}^{a}{}_{k]} = 0
	\quad \Ra \quad 
	\tau(\hat\Sigma)_{ijk}{}^a = 0 .
\end{equation}
In particular, we observe that the vanishing of this component of the intrinsic torsion is simply equivalent to the algebra closure condition that the KD derivative of a Killing spinor is again a Killing spinor.

We conclude that for any supersymmetric background the entire generalised intrinsic torsion must vanish. We can therefore find a $\gN$ compatible $\hat\Dgen$ which is torsion-free if the supersymmetry equations~\eqref{eq:KSE7} hold. 

Since, conversely, if the generalised intrinsic torsion is zero, the Killing spinor equations are satisfied trivially, we have a precise equivalence between supersymmetric backgrounds and generalised $\gN$ special holonomy spaces. Indeed, we have found an isomorphism
\begin{equation}
   \Tint \simeq \mathcal{N}\times(S \oplus J) \oplus \tbinom{\mathcal{N}}{3} \times V\oplus\text{c.c.} ,
\end{equation}
where $V$ is the bundle associated to the $\rep{[8-\mathcal{N}]}$ representation and which is constrained by $(\LgenS_{V_{ij}} \epsilon_k)^a$.

We stress again that while we have focused on the $d=7$ case, the exact same proof holds in lower dimensions as well, with the structure groups listed in table~\ref{table2}.


\section{Discussion and outlook}
\label{sec:outlook}

We have finally been able to answer the question of whether generic supersymmetric backgrounds may be described purely in terms of geometric integrability conditions. Indeed, they are exactly torsion-free structures on the generalised tangent bundle. We have shown that at each level $\mathcal{N}$ of preserved supersymmetry, there exists a single generalised $\gN$ structure with vanishing intrinsic torsion and, conversely, every space admitting a torsion-free $\gN$ structure is a solution of $\mathcal{N}$ independent Killing spinor equations. (However, even though generalised special holonomy guarantees that equations of motion are solved since these spaces are necessarily generalised Ricci-flat, it is important to keep in mind that in order to actually build genuine global Minkowski compactifications, it will still be necessary to address the no-go theorems~\cite{Candelas:1984yd,Maldacena:2000mw,Ivanov:2000fg}.)

We have also introduced a new tool to generalised geometry, the Kosmann-Dorfman derivative, which might prove useful in further applications of the formalism. In particular, it allowed us to give a description of the Killing superalgebra that arises in the internal space of supersymmetric backgrounds. Clearly, all the crucial closure and Jacobi properties of this algebra become trivial once it is shown that these spaces have generalised special holonomy -- just as the usual special holonomy manifolds trivially satisfy the requirements for a fluxless Killing superalgebra. It nonetheless revealed the perhaps underappreciated fact that even in the presence of fluxes the internal spaces of supersymmetric Minkowski compactifications admit a large number of commuting isometries which preserve the Killing spinors. Of course, we have not shown that these are generally independent or even non-vanishing. In order to determine the precise number of geometric isometries arising in this way, one would have to examine the vector components of the Killing spinor bilinears, which can be done only by analysing the possible orbits of multiple spinors at $\Spin(d) \subset \dHd$ level. However, this could provide very strong constraints on $\mathcal{N}\geq3$ solutions, and may explain why so few examples are known.\footnote{A rare example of an $\mathcal{N}=3$ Minkowski compactification was given in~\cite{Frey:2002hf}.} Note that, in fact, for $\mathcal{N}\geq5$ this analysis has effectively already been done in~\cite{FigueroaO'Farrill:2012fp}, which proves that all such solutions are homogeneous, so that the warp-factor equation of motion implies that all fluxes vanish and the geometry is flat. 

Another observation made apparent by these commutation relations is that the eleven-dimensional Killing superalgebra of such backgrounds is always the supertranslational part of the usual $\mathcal{N}$-extended super-Poincar\'e algebra. This result could have been arrived at by algebraic means if one is willing to make some physical assumptions~\cite{Haag:1974qh,Sohnius:1985qm}, however, we have shown that it follows purely by geometrical means starting from eleven-dimensional supergravity. Also, in this construction the central charges gain a neat realisation as the commuting generalised Killing vectors on the internal space.  One would expect that the algebra of these would give some parts of the embedding tensor~\cite{deWit:2002vt} of a consistent truncation of eleven-dimensional supergravity around such Minkowski backgrounds. Our results would therefore seem to give constraints on possible embedding tensors. 

On the way to our main conclusions we made use of the lemma that, in the non-conformal split frame, the (Kosmann-)Dorfman derivative along a generalised Killing vector reduces to the Lie derivative along its vector part. This was previously noted and used to show that the generalised Reeb vector gives rise to the R-charges of various objects in the constructions of~\cite{Ashmore:2016qvs,Ashmore:2016oug,Grana:2016dyl}. 
This equality is extremely useful in proving relations such as~\eqref{eq:Kosmann-closure} and~\eqref{eq:gen-comm-susy-lie}, as when written out in terms of ordinary geometry objects these relations become transparent. However, it is much more difficult to establish these relations without using this decomposition (i.e. the anchor map). Working only in generalised geometry objects with $\dHd$ symmetry, one needs to do substantial manipulations to show cancellations involving those combinations of second partial derivatives which vanish identically (see~\cite{CSW2}). Even then, one needs to do further manipulations to show that certain contributions of the fluxes to the torsion of the generalised connection also cancel appropriately. This seems to require a quadratic constraint on the torsion much like the quadratic constraint satisfied by the embedding tensor of gauged supergravity~\cite{deWit:2002vt}. These objects transform in the same representation of the exceptional group and moreover at a point the generalised torsion takes the same form as the embedding tensor of an ordinary Scherk-Schwarz reduction with fluxes (see~\cite{LSW1}, appendix C). Therefore, one expects that they will satisfy the same quadratic relation and that it is this which gives the cancellations required for~\eqref{eq:Kosmann-closure} and~\eqref{eq:gen-comm-susy-lie}. Again, via the definition of the generalised torsion~\cite{CSW2}, these constraints are linked back to the parabolic nature of the Dorfman bracket and its Leibniz property.

An important remark at this point is that, since the generalised Killing vector resulting from the bracket of two Killing spinors includes all the spinor bilinears, the Kosmann-Dorfman derivative along such a generalised vector defines an action of these $p$-forms on Killing spinors. Furthermore, we see that the action of such $p$-form bilinears should be considered simultaneously in precisely the combination fixed by the $\dHd$-covariant form of the bracket in order to obtain the correct KD algebraic structure. On the other hand, the lemma~\eqref{eq:spinor-Dorfman-GKV} implies that in an appropriate frame this reduces to the action of just the vector bilinear, as the Killing spinor equations force the terms involving $p$-form bilinears to be related to the fluxes. It would therefore be interesting to relate this construction to the difficulties raised in~\cite{FigueroaO'Farrill:2008ka} regarding the extension of the brackets of the Killing superalgebra to accommodate ``supergravity Killing forms''. There such a bracket is given on a very restricted set of spacetimes (more recent work on this has appeared in~\cite{Ertem:2016fso}, though again working only on a special class of spacetimes). While the Kosmann-Dorfman bracket requires no such assumptions, it is defined only on the $d<8$ dimensional internal space. It thus remains an open question whether this construction can be extended to eleven-dimensions.

Having solved the problem of assigning a geometric integrability condition to fully generic Minkowski backgrounds, it is natural to consider expanding these methods to other supersymmetric backgrounds. In particular, it should be possible to extend the results of~\cite{CS1} for $\mathcal{N}=1$ AdS to arbitrary amounts of preserved supersymmetry. There are some minor technical complexities associated with AdS backgrounds which should require some statements to be modified slightly, and we hope to investigate this further in the near future.

Another potential question concerns the possible definition of generalised holonomy for the exceptional geometry relevant to massive type IIA supergravity~\cite{Cassani:2016ncu}. There, the generalised tangent space is isomorphic to that for massless type IIA and the generalised $\gN$ structures would be the same for each level of supersymmetry. However, due to the different bracket structure, the notion of integrability of these structures would be subtly different. In fact there would be a one-parameter family of integrability conditions, vaguely similar in nature to the weak holonomy conditions of~\cite{CS1}. It would therefore be interesting to see how exactly this would work out. One could similarly investigate the situation in the heterotic theory using the geometries of~\cite{Garcia-Fernandez:2013gja,CMTW}.

Finally, one of the most compelling reasons to wish to rewrite the supersymmetry conditions as torsion-free structures is the study of their moduli spaces. Already these integrability results have enabled significant progress in this direction for the H- and V-structures associated to eight supercharge vacua~\cite{Ashmore:2015joa,Ashmore:2016qvs,Ashmore:2016oug} (see also~\cite{Garcia-Fernandez:2015hja} for a generalised geometric construction of moduli for the Strominger system in the heterotic case). One would expect that the integrability conditions we have established here could be used to describe the moduli for general backgrounds with $\mathcal{N}\geq3$ supersymmetry, for which the moduli space itself is fixed by supersymmetry (see e.g.~\cite{Freedman:2012zz}), corroborating and extending the results of the earlier study~\cite{Triendl:2009ap}.


\begin{acknowledgments}
We would like to thank Jos\'e Figueroa-O'Farrill, Mariana Gra\~na, George Papadopoulos, and especially Dan Waldram for helpful discussions.
A.~C.~has been supported by the Laboratoire d'Excellence CARMIN. 
C.~S-C.~has been supported by the ANR grant 12-BF05-003-002 and a grant from the Foundational Questions Institute (FQXi) Fund, a donor advised fund of the Silicon Valley Community Foundation on the basis of proposal FQXi-RFP3-1321 (this grant was administered by Theiss Research).
C.~S-C.~thanks the CERN-CKC TH Institute on Duality Symmetries in String and M-Theories for hospitality.
A.~C.~and C.~S-C.~also thank the Program on the Mathematics of String Theory at Institut Henri Poincar\'e, Paris for hospitality during the completion of this work.
\end{acknowledgments}


\appendix


\section{Evaluation of the Kosmann-Dorfman derivative by a GKV}
\label{app:spinor-Dorfman-GKV}

In this appendix we provide some details of the calculation which leads to the useful lemma~\eqref{eq:spinor-Dorfman-GKV}, which states that the Kosmann-Dorfman derivative of a spinor field by a GKV is equal to the ordinary Lie derivative of the spinor field along the vector part of the GKV.

We study the seven-dimensional case, working under an $\SO(8)$ decomposition of $\SU(8)$ (as in section 5 of~\cite{CSW3}). However, this is sufficient to prove the result in general dimension, as one could have done the same calculation in the dimension-independent formalism for fermions (as in section 4 of~\cite{CSW3}) and one must get the same answer.

Recall that, using $SO(8)$ gamma matrices $\hat\gamma$, one can express a torsion-free $\SU(8)$ compatible connection in the form~\cite{CSW3}
\begin{equation}
\label{eq:SO8-derivative}
\begin{aligned}
   \Dgen_{ii'} \chi &= \Dgen^\nabla_{ii'} \chi \,
      + \tfrac14 \Sigma_{ii'jj'} \gamh^{jj'} \chi
      - \tfrac{1}{48}\ii \Sigma_{ii' k_1 \dots k_4} 
         \gamh^{k_1 \dots k_4} \chi ,\\
   \tilde\Dgen_{ii'} \chi &= \tilde\Dgen^\nabla_{ii'} \chi
      + \tfrac14 \tilde{\Sigma}_{ii'jj'} \gamh^{jj'} \chi
      - \tfrac{1}{48}\ii \tilde{\Sigma}_{ii'k_1 \dots k_4} 
         \gamh^{k_1 \dots k_4} \chi ,
\end{aligned}
\end{equation}
where, if one expresses the components of the generalised vector with respect to the conformal split frame, 
\begin{equation}
\label{eq:conection-SO8}
\begin{aligned}
   \Sigma_{ii'jj'} &= -\tfrac13\ee^{\Delta} \delta_{ij}\tilde{K}_{i'j'} 
      + \tfrac{1}{42}\ee^{\Delta}\tF \delta_{ij}\delta_{i'j'}
      - 
      \delta_{ij}\der_{i'j'}\Delta + \am_{ii'jj'}, \\
   \tilde{\Sigma}_{ii'jj'} &= \tfrac13 \ee^{\Delta}K_{ii'jj'}
      - \tfrac16 \ee^{\Delta}K_{jj'ii'} + \tilde{\am}_{ii'jj'}, \\
   \Sigma_{i_1\dots i_6} &= \am_{i_1\dots i_6}, \\
   \tilde{\Sigma}_{i_1\dots i_6} &= \tilde{\am}_{i_1\dots i_6} .
\end{aligned}
\end{equation}
Here, primed and unprimed indices are antisymmetrised
implicitly and $(\am,\tilde{\am})$ are the parts of the connection which are not determined
by the condition that the connection be torsion-free and compatible. The supergravity fluxes enter this expression as $\tF = \tfrac{1}{7!} \epsilon^{a_1 \dots a_7} \tF_{a_1 \dots a_7}$ and 
\begin{equation}
\label{eq:SO7-flux}
\begin{aligned}
   K_{ii'jj'} &= \begin{cases}
         (*F)_{abc} \, &\text{for } (i,i,'j,j')=(a,b,c,8) \\
         0 \, &\text{otherwise} 
      \end{cases}, \\
   \tilde{K}_{ij} &= \begin{cases} 
         \tF \, &\text{for } (i,j)=(8,8) \\
         0 \, &\text{otherwise} 
      \end{cases},
\end{aligned}
\end{equation}
and the partial derivative and Levi-Civita connection are written with $\SO(8)$ indices as
\begin{equation}
\label{eq:SO7-derivatives}
\begin{aligned}
   \der_{a8\ph{'}} &= \tfrac12 \ee^\Delta \der_a , && \qquad \qquad &&&
   \der_{ab\ph{'}} &= 0 , && \qquad \qquad &&&
   \tilde\der^{ii'} &= 0 , \\
   \Dgen^\nabla_{a8\ph{'}} &= \tfrac12 \ee^\Delta \LC_a , && \qquad \qquad &&&
   \Dgen^\nabla_{ab\ph{'}} &= 0 , && \qquad \qquad &&&
   \tilde\Dgen^\nabla{}^{ii'} &= 0 .
\end{aligned}
\end{equation}
The relations we need between $\SO(8)$ and $\SU(8)$ indices are then 
\begin{equation}
\begin{aligned}
   V^{\alpha \beta} &=  \ii (\gamh_{ij})^{\alpha \beta} 
      \big(V^{ij} + \ii \tV^{ij }\big) ,
      &\hs{30pt} & & \Dgen^{\alpha \beta} &=  \ii (\gamh^{ij})^{\alpha \beta} 
      \big( \Dgen_{ij} + \ii \tilde\Dgen_{ij }\big) ,\\
   \bar{V}_{\alpha \beta} &= - \ii (\gamh^{ij})_{\alpha \beta} 
      \big( V_{ij} - \ii \tV_{ij }\big) ,
      && &\bar\Dgen_{\alpha \beta} &= - \ii (\gamh_{ij})_{\alpha \beta} 
      \big( \Dgen^{ij} - \ii \tilde\Dgen^{ij }\big).
\end{aligned}
\end{equation}

Now keeping in mind the very useful completeness relations
\begin{align*}
   \gamh^{ij}{}_{\alpha \beta} \gamh_{ij}{}^{\gamma \delta} 
      &= 16 \delta^{\gamma \delta}_{\alpha \beta} , &
   \gamh^{ij}{}_{\alpha \beta} \gamh_{kl}{}^{\alpha \beta} 
      &= 16 \delta^{ij}_{kl} ,
\end{align*}
we are ready to substitute these expressions into the formula~\eqref{eq:spinor-Dorfman-su8} for the Kosmann-Dorfman derivative in $\SU(8)$ indices. An initial intermediate step reached is the $\SO(8)$ form of the KD derivative
\begin{equation}
\begin{aligned}
	\LgenS_V \epsilon
	& = ( V^{ij} \Dgen_{ij} + \tV^{ij} \tilde\Dgen_{ij} ) \epsilon
		+ \tfrac12 (\Dgen_{ik} V_j{}^k + \tilde\Dgen_{ik} \tV_j{}^k) \gamh^{ij} \epsilon 
		+ \tfrac{\ii}{8} (\Dgen_{ij} \tV_{kl} - \tilde\Dgen_{ij} V_{kl}) \gamma^{ijkl} \epsilon .
\end{aligned}
\end{equation}
We can then substitute in~\eqref{eq:SO8-derivative} and~\eqref{eq:conection-SO8}. Note that we can immediately disregard the terms $(\am,\tilde{\am})$ from equation~\eqref{eq:conection-SO8} as these will necessarily cancel out of the final answer since the KD derivative depends only on torsion components. We obtain an expression in terms of fluxes embedded in $\SO(8)$ representations. Decomposing under $\SO(7)$ with~\eqref{eq:SO7-flux} and~\eqref{eq:SO7-derivatives}, one arrives at
\begin{equation}
\begin{aligned}
	\LgenS_V \epsilon
	= \ee^{\Delta} \Big( v^a \LC_a \epsilon &+ \tfrac14 \LC_{a} v_{b} \gamma^{ab} \epsilon
		+ \tfrac14  (\der_{a} \Delta) \, v_{b} \gamma^{ab} \epsilon \\
		&+ \tfrac14 \tfrac1{3!} [ \dd \omega + \dd \Delta \wedge \omega
			- i_v F]_{abc} \gamma^{abc} \epsilon \\
		&+ \tfrac14 \tfrac1{6!} [ \dd \sigma + \dd \Delta \wedge \sigma 
			- i_v \tF + \omega \wedge F]_{a_1 \dots a_6} \gamma^{a_1 \dots a_6} \epsilon 
			\Big) .
\end{aligned}
\end{equation}
The terms involving the fluxes $F$ and $\tF$ cancel due to the GKV condition~\eqref{eq:GKV-non-conf} (evaluated in a conformal split frame) and one has simply 
\begin{equation}
\begin{aligned}
	\LgenS_V \epsilon
	& = \ee^{\Delta} \Big( v^a \LC_a \epsilon + \tfrac14 \LC_{a} v_{b} \gamma^{ab} \epsilon
		+ \tfrac14  (\der_{a} \Delta) \, v_{b} \gamma^{ab} \epsilon \Big) \\
	&= \mathcal{L}_{\ee^\Delta v} \; \epsilon .
\end{aligned}
\end{equation}
Here we recognise $\ee^{\Delta} v$ as the vector component of $V$ in the non-conformal split frame, so we have arrived at~\eqref{eq:spinor-Dorfman-GKV}. Note that, since the KD bracket is Leibniz, and since when acting on arbitrary generalised vectors along GKVs it matches the Dorfman derivative~\eqref{eq:Dorfman-GKV}, we have also just proven~\eqref{eq:Dorfman-GKV-Lv}.


\section{Decompositions and tensor products}

We list the representations associated to the generalised tensor bundles from section~\ref{sec:gen-sp-hol} and their decompositions under the reduced structure groups imposed by supersymmetry.
 

\subsection{The torsion space $W$}
\label{app:torsion}

In~\cite{CSW2} it was shown that the generalised torsion $T(D)$ of a generalised, metric-compatible connection in $E_{7(7)}\times\bbR^+$ generalised geometry takes values in a bundle $W$ with fibres transforming in the $\rep{28}+\rep{36}+\rep{420}+\CC$ of $\SU(8)$.

These representations can then be decomposed under the reduced structure group $\SU(8-\mathcal{N})\subset \SU(8)$ as follows
\begin{itemize}
\item $\mathcal{N}=1,$ $ \SU(7)$
\begin{equation}
\begin{aligned}[]
	\rep{28}+\rep{36}+\rep{420}+\CC \ra (\rep{7}+\rep{21})+(\rep{1}+\rep{7}+\rep{28})+(\rep{21} + \rep{35} + \rep{140} + \rep{224})+\CC
\end{aligned}
\end{equation}
\item $\mathcal{N}=2,$ $ \SU(6)$
\begin{equation}
\begin{aligned}[]
	\rep{28}+\rep{36}+\rep{420}+\CC \ra &(\rep{1}+2\times\rep{6}+\rep{15})+(3\times\rep{1}+2\times\rep{6}+\rep{21})\\&+ (2\times\rep{6} + 4\times\rep{15} + 2\times\rep{20} + \rep{35} + 2\times\rep{84} + \rep{105})+\CC
\end{aligned}
\end{equation}
\item $\mathcal{N}=3,$ $ \SU(5)$
\begin{equation}
\begin{aligned}[]
	\rep{28}+\rep{36}+\rep{420}+\CC \ra &(3\times\rep{1}+3\times\rep{5}+\rep{10})+(6\times\rep{1}+3\times\rep{5}+\rep{15})\\&+ (3\times\rep{1} + 10\times\rep{5} 
		  + 12\times\rep{10}    + 3\times\rep{24} 
		+ \rep{40} +3\times \rep{45})+\CC
\end{aligned}
\end{equation}
\item $\mathcal{N}=4,$ $ \SU(4)$
\begin{equation}
\begin{aligned}[]
	\rep{28}+\rep{36}+\rep{420} +\CC\ra &(6\times\rep{1}+4\times\rep{4}+\rep{6})+(10\times\rep{1}+4\times\rep{4}+\rep{10}) \\
	&+( 16\times\rep{1} + 32\times\rep{4} 
		+ 16\times\rep{6} + \rep{10} 
		  + 6\times\rep{15}
		+ 4\times\rep{20}  )+\CC
\end{aligned}
\end{equation}
\item $\mathcal{N}=5,$ $ \SU(3)$
\begin{equation}
\begin{aligned}[]
	\rep{28}+\rep{36}+\rep{420} +\CC\ra &(10\times\rep{1}+6\times\rep{3})+(15\times\rep{1}+5\times\rep{3}+\rep{6})\\&+ (55\times\rep{1} + 85\times\rep{3} 
		+ 5\times\rep{6}   + 10\times\rep{8})+\CC
\end{aligned}
\end{equation}
\item $\mathcal{N}=6,$ $ \SU(2)$
\begin{equation}
\begin{aligned}[]
	\rep{28}+\rep{36}+\rep{420}+\CC \ra &(16\times\rep{1}+6\times\rep{2})+(21\times\rep{1}+6\times\rep{2}+\rep{3})\\&+ (155\times\rep{1}  + 110\times\rep{2}  + 15\times\rep{3}  )+\CC
\end{aligned}
\end{equation}
\end{itemize}
%


\subsection{The space of compatible connections $\Kgen_{\SU(8-\mathcal{N})}$}
\label{app:comp-c}

The difference between any two $SU(8-\mathcal{N})$ compatible connections is given by an element of
\begin{equation}
	 \Kgen_{\SU(8-\mathcal{N})} = E^* \otimes \adj P_{\SU(8-\mathcal{N})} .
\end{equation}
We list the corresponding $\SU(8-\mathcal{N})$ representations for these tensor products for each value of $\mathcal{N}$.
\begin{itemize}
\item $\mathcal{N}=1,$ $ \SU(7)$
\begin{equation}
\begin{aligned}[]
	K_{\SU(7)} &=  (\rep{7} + \rep{21}) \times \rep{48}  + \text{c.c.} \\
	&=   \rep{7} + \rep{21} + \rep{28} + \rep{140} + \rep{189} + \rep{224} + \rep{735} 
		+  \text{c.c.} 
\end{aligned}
\end{equation}
\item $\mathcal{N}=2,$ $ \SU(6)$
\begin{equation}
\begin{aligned}[]
	K_{\SU(6)} &=
		 ( \rep{1} + 2 \times \rep{6} + \rep{15}) \times \rep{35}   +  \text{c.c.}  \\
	 & =  2 \times \rep{6} + \rep{15} + \rep{21}+ \rep{35} + 2 \times \rep{84} + \rep{105}
	 + 2 \times \rep{120} + \rep{384} +  \text{c.c.} 
\end{aligned}
\end{equation}
\item $\mathcal{N}=3,$ $ \SU(5)$
\begin{equation}
\begin{aligned}[]
	K_{\SU(5)} &=  (3\times \rep{1} + 3\times \rep{5} + \rep{10} ) \times \rep{24}  
		+ \text{c.c.}  \\
		&= 3 \times \rep{5} + \rep{10} + \rep{15} + 3 \times \rep{24} + 2 \times \rep{45} 
			+ 4 \times \rep{70} + \rep{175}   +  \text{c.c.}  
\end{aligned}
\end{equation}
\item $\mathcal{N}=4,$ $ \SU(4)$
\begin{equation}
\begin{aligned}[]
	K_{\SU(4)} &= (6\times \rep{1} + 4\times \rep{4} + \rep{6} ) \times \rep{15}  
		+  \text{c.c.}  \\
		&=  4 \times \rep{4} + \rep{6} + 2\times\rep{10} + 6 \times \rep{15} + 4 \times \rep{20} 
			+ 4 \times \rep{36} + \rep{64}   +  \text{c.c.}
\end{aligned}
\end{equation}
\item $\mathcal{N}=5,$ $ \SU(3)$
\begin{equation}
\begin{aligned}[]
	K_{\SU(3)} &=(10\times \rep{1} + 5\times \rep{3} + \rep{\bar3} ) \times \rep{8}  
		+ \text{c.c.}  \\
		&= 6 \times \rep{3} + 6\times\rep{6} + 10\times\rep{8} + 6 \times \rep{15}  
			+ \text{c.c.} 
\end{aligned}
\end{equation}
\item $\mathcal{N}=6,$ $ \SU(2)$
\begin{equation}
\begin{aligned}[]
	K_{\SU(2)} = (32\times \rep{1} + 12\times \rep{2} ) \times \rep{3}  
		= 12 \times \rep{2} + 32\times\rep{3} + 12 \times \rep{4}
\end{aligned}
\end{equation}
\end{itemize}
%


\subsection{The intrinsic torsion space $\Tint$}
\label{app:int-t}

Below, we list the $\SU(8-\mathcal{N})$ representations appearing in the intrinsic torsion of the reduced structure $P_{\SU(8-\mathcal{N})}$ for $\mathcal{N} = 1, \dots , 6$.
\begin{itemize}
\item $\mathcal{N}=1,$ $ \SU(7)$
\begin{equation}
\begin{aligned}[]
	\Tint(P_{\SU(7)})
	= \rep{1} + \rep{7} + \rep{21} + \rep{35} 
		+  \text{c.c.} 
\end{aligned}
\end{equation}
\item $\mathcal{N}=2,$ $ \SU(6)$
\begin{equation}
\begin{aligned}[]
	\Tint(P_{\SU(6)})
	&=  4 \times \rep{1} + 4\times\rep{6} + 4\times\rep{15} + 2\times\rep{20} 
		+  \text{c.c.} \\
		&=  2 \times [ 2 \times \rep{1} + 2\times\rep{6} + 2\times\rep{15} + \rep{20}] 
			+ \text{c.c.} 
\end{aligned}
\end{equation}
\item $\mathcal{N}=3,$ $ \SU(5)$
\begin{equation}
\begin{aligned}[]
	\Tint(P_{\SU(5)})
	&=  12 \times \rep{1} + 13\times\rep{5} + 12\times\rep{10} +  \text{c.c.}  \\
		&= 3 \times [ 4 \times \rep{1} + 4\times\rep{5} + 4\times\rep{10}] 
		+ \tbinom{3}{3}\times\rep{5} 
			+ \text{c.c.}  
\end{aligned}
\end{equation}
\item $\mathcal{N}=4,$ $ \SU(4)$
\begin{equation}
\begin{aligned}[]
	\Tint(P_{\SU(4)})
	&=  32 \times \rep{1} + 36\times\rep{4} + 16\times\rep{6}  +  \text{c.c.}  \\
		&= 4 \times [ 8 \times \rep{1} + 8\times\rep{4} + 4\times\rep{6}] 
		+ \tbinom{4}{3}\times\rep{4}
			+  \text{c.c.} 
\end{aligned}
\end{equation}
\item $\mathcal{N}=5,$ $ \SU(3)$
\begin{equation}
\begin{aligned}[]
	\Tint(P_{\SU(3)}) 
	&=  80 \times \rep{1} + 90\times\rep{3}  +  \text{c.c.}  \\
		&= 5 \times [ 16 \times \rep{1} + 16\times\rep{3} ] + \tbinom{5}{3}\times\rep{3} 
			+  \text{c.c.} 
\end{aligned}
\end{equation}
\item $\mathcal{N}=6,$ $\SU(2)$
\begin{equation}
\begin{aligned}[]
	\Tint(P_{\SU(2)}) 
		&= 192 \times \rep{1} + 116\times\rep{2} +  \text{c.c.}  \\
		&= 6 \times [ 32\times\rep{1} + 16 \times \rep{2} ] + \tbinom{6}{3} \times \rep{2} 
			+  \text{c.c.} 
\end{aligned}
\end{equation}
\end{itemize}
%


\section{Explicit calculation of the intrinsic torsion}

In this appendix we explore in more detail the computation of the explicit projections which give the generalised intrinsic torsion of an $SU(8-\mathcal{N})$ generalised connection. We present two different ways of obtaining the result. 


\subsection{Alternative computation in terms of reducible representations}
\label{app:int-torsion-calc}

Here we will give a modified version of the intrinsic torsion computation of section~\ref{sec:gen-hol} which, while still in terms of $\gN$ reducible objects, is perhaps the cleanest way to reproduce the results in other dimensions for the interested reader.
Note that all expressions in this appendix should be understood to also be accompanied by their complex conjugates, though for the sake of clarity we will not write them explicitly.

First, we define a map
\begin{equation}
\label{eq:tauhat}
\begin{aligned}
	\hat\tau : \HConSp &\ra \hat{W} = (\rep{28} + \rep{36}) + (\rep{28} + \rep{420}) ,\\
		\hat\Sigma &\mapsto ( \hat\tau_{\alpha\beta}, \hat\tau_{\alpha\beta}{}^{\gamma}{}_\delta) 
			=  (\hat\Sigma_{\alpha\gamma}{}^{\gamma}{}_\beta , 
				\hat\Sigma_{[\alpha\beta}{}^{\gamma}{}_{\delta]}),
\end{aligned}
\end{equation}
and also the projection
\begin{equation}
\label{eq:Im-tau-projector}
\begin{aligned}
	p_0 : \hat{W} &\ra \rep{28} ,\\
		( \hat\tau_{\alpha\beta}, \hat\tau_{\alpha\beta}{}^{\gamma}{}_\delta)   &\mapsto 
			(\hat\tau_{\alpha\beta}{}^{\gamma}{}_\gamma + \tfrac23 \hat\tau_{[\alpha\beta]}),
\end{aligned}
\end{equation}
which has $\text{Im}\, \hat\tau = \text{ker}\, p_0$ so that the following sequence is exact
\begin{equation}
	\HConSp \stackrel{\hat\tau}{\longrightarrow} \hat{W} \stackrel{p_0}{\longrightarrow} \rep{28}.
\end{equation}
Next we define
\begin{equation}
\label{eq:hatW-W-projector}
\begin{aligned}
	p_1 : \hat{W} &\ra W ,\\
		( \hat\tau_{\alpha\beta}, \hat\tau_{\alpha\beta}{}^{\gamma}{}_\delta)  &\mapsto 
			(\hat\tau_{\alpha\beta}, \,
			\hat\tau_{\alpha\beta}{}^{\gamma}{}_\delta 
				+ \tfrac13 \hat\tau_{[\alpha\beta} \delta^\gamma{}_{\delta]}) ,
\end{aligned}
\end{equation}
which projects the reducible tensor $\hat\tau_{\alpha\beta}{}^{\gamma}{}_\delta$ onto its traceless part in the $\rep{420}$ representation. Its kernel is thus
\begin{equation}
	\text{ker}\, p_1 = \{ (0 , \kappa_{[\alpha\beta} \delta^\gamma{}_{\delta]}) \in \hat{W} \} ,
\end{equation}
and we have that our map $\tau$ from~\eqref{eq:tau} is
\begin{equation}
\label{eq:tau-hat-tau}
	\tau = p_1 \circ \hat\tau : \HConSp \ra W .
\end{equation}
Finally, we note that as $\text{ker}\, p_0 \cap \text{ker}\, p_1 = 0$, we have that $p_1$ restricted to $\text{ker}\, p_0 = \text{Im}\, \hat\tau$ is an isomorphism.

Next we restrict to the space of compatible connections $\Kgen_{\gN}$, defining $\hat\tau_{\gN} = \hat\tau|_{\Kgen_{\gN}}$.
The intrinsic torsion was defined in section~\ref{sec:gen-sp-hol} as the projection of the torsion onto $W_{\text{int}} = W / W_{\gN} = \text{Im}\, \tau / \text{Im}\, \tau_{\gN}$.  The key point in considering the reducible objects in $\hat{W}$ is that, as $p_1 |_{\text{Im}\, \hat\tau}$ is an isomorphism, we have that $p_1$ also induces an isomorphism
\begin{equation}
\label{eq:int-torsion-isom}
\begin{aligned}
	\frac{\text{Im}\, \hat\tau}{\text{Im}\, \hat\tau_{\gN}} \stackrel{p_1|}{\longrightarrow}
	 \frac{\text{Im}\, \tau}{\text{Im}\, \tau_{\gN}} = \frac{W}{W_{\gN}} .
\end{aligned}
\end{equation}

The intrinsic torsion is composed of those irreducible parts of $W$ which are zero for any compatible connection, i.e. the cokernel of $\tau_{\gN}$. By~\eqref{eq:tau-hat-tau} and~\eqref{eq:int-torsion-isom}, we can equally well focus on which parts of the image of $\hat\tau$ are identically zero. If $\Sigma$ is an $\SU(8-\mathcal{N})$ compatible connection, i.e. $\Sigma^{\alpha\beta\gamma}{}_i = \Sigma^{\alpha\beta i}{}_\gamma = \Sigma^{\alpha\beta c}{}_c = 0$, then we find 
\begin{equation}
\label{eq:map3}
\begin{aligned}
   \hat\tau(\Sigma)_{ab} &= \Sigma_{ac}{}^{c}{}_b  , & 
      && && && 
   \hat\tau(\Sigma)_{ai} &= 0   , \\
   \hat\tau(\Sigma)_{ia} &= \Sigma_{ic}{}^{c}{}_a , & 
      && && && 
   \hat\tau(\Sigma)_{ij}  &=0, \\
   \hat\tau(\Sigma)_{ab}{}^{c}{}_d &= \Sigma_{[ab}{}^{c}{}_{d]}
       , & 
      && && && 
   \hat\tau(\Sigma)_{ab}{}^{c}{}_i &= 0  , \\
   \hat\tau(\Sigma)_{ia}{}^{b}{}_c &= \tfrac23 \Sigma_{i[a}{}^{b}{}_{c]}
      ,& 
      && && && 
   \hat\tau(\Sigma)_{ab}{}^{i}{}_j &= 0   , \\
   \hat\tau(\Sigma)_{ij}{}^{a}{}_b & = \tfrac13 \Sigma_{ij}{}^{a}{}_b
      ,&
      && && && 
   \hat\tau(\Sigma)_{ai}{}^{j}{}_k &= 0   , \\ 
    \phantom{ \hat\tau(\Sigma)_{ij}{}^a{}_{k} }&\phantom{= 0
       ,} &
      && && && 
   \hat\tau(\Sigma)_{ij}{}^{k}{}_l &= 0 , \\ 
   \phantom{ \hat\tau(\Sigma)_{ij}{}^{a}{}^k }&\phantom{= 0
       ,} &
      && && && 
    \hat\tau(\Sigma)_{ij}{}^{a}{}_k &= 0 .
\end{aligned}
\end{equation}
We thus see by inspection that the parts of $\text{Im}\, \hat\tau$ in the left-hand column can generically be non-zero and are also independent up to satisfying $p_0 (\hat\tau(\Sigma))=0$, which is automatic as they are in $\text{Im}\, \hat\tau$. Thus, they effectively give a basis for the image of $\hat\tau_{\gN}$. The parts in the right-hand column vanish and are similarly independent, thus these can also be viewed as giving a basis for $\text{Im}\, \hat\tau / \text{Im}\, \hat\tau_{\gN}$. The column on the right then gives the the intrinsic torsion -- no matter our choice of $\Sigma$, the torsion of any connection lying in those components cannot be shifted since there the $\hat\tau$ map acting on $\Sigma$ vanishes. Conversely, if all of the components in the right column do vanish, then we can set the torsion to zero by shifting by a compatible $\Sigma$, due to the independence of the components in the left column. We conclude that the intrinsic torsion is parameterised by the quantities
\begin{equation}
\label{eq:int-torsion-red2}
\begin{aligned}
	\hat\tau_{\text{int}} = ( \hat\tau_{ai} , \hat\tau_{ij}, \hat\tau_{ab}{}^{i}{}_c, \hat\tau_{ab}{}^{i}{}_j, \hat\tau_{ai}{}^{j}{}_k, 
		\hat\tau_{ij}{}^{k}{}_l, \hat\tau_{ij}{}^{a}{}_k) ,
\end{aligned}
\end{equation}
which are again automatically constrained so that $\hat\tau_{\text{int}}$ lies in the kernel of $p_0$.


\subsection{Alternative computation in terms of irreducible representations}
\label{app:int-torsion-irrep}

Finally, we will now give another alternative method of obtaining the intrinsic torsion conditions, now working exclusively with irreps. Given the torsion map~\eqref{eq:torsion-su8}, let us define
\begin{equation}
\label{eq:torsion-28-36-420}
\begin{aligned}
   \tau(\hat\Sigma)_{\alpha\beta} 
      &= \hat\Sigma_{\alpha\gamma}{}^\gamma{}_{\beta} 
      = -\tfrac23 \alpha_{[\alpha\beta]} + \beta_{(\alpha\beta)}
      && \in   \rep{28} + \rep{36} , \\
   \tau(\hat\Sigma)_{\alpha\beta}{}_\delta {}^{\gamma}
      &= \hat\Sigma_{[\alpha\beta}{}^{\gamma}{}_{\delta]} 
      	- \tfrac12 \alpha_{[\alpha\beta} \delta^{\gamma}{}_{\delta]}
      && \in \rep{420} , 
\end{aligned}
\end{equation}
together with their conjugates. We again write the indices as $\alpha = (a,i)$ and decompose into irreducible $\SU(8-\mathcal{N})\times\SU(\mathcal{N})$ parts. We thus define
\begin{equation}
\label{eq:420-decomp}
\begin{aligned}
	\tau_{ab}{}_d{}^{c} &= \Xa{}_{ab}{}^{c}{}_{d} + \tfrac{3}{6-\mathcal{N}} \Ya{}_{[ab}\delta^{c}{}_{d]} ,  \\
	\tau_{ij}{}_l{}^{k} &= \Xb{}_{ij}{}^{k}{}_{l} + \tfrac{3}{\mathcal{N}-2} \Yb{}_{[ij}\delta^{k}{}_{l]} ,  \\
	\tau_{ia}{}_c{}^{b} &= \Xc{}_{ia}{}^{b}{}_{c} + \tfrac{2}{7-\mathcal{N}} \Yc{}_{i[a}\delta^{b}{}_{c]} ,  \\
	\tau_{ai}{}_k{}^{j} &= \Xd{}_{ai}{}^{j}{}_{k} + \tfrac{2}{\mathcal{N}-1} \Yc{}_{a[i}\delta^{j}{}_{k]} ,  \\
	\tau_{ij}{}_b{}^{a} &= \Xe{}_{ij}{}^{a}{}_{b} - \tfrac{1}{8-\mathcal{N}} \Yb{}_{ij}\delta^{a}{}_{b} ,  \\
	\tau_{ab}{}_j{}^{i} &= \Xf{}_{ab}{}^{i}{}_{j} - \tfrac{1}{\mathcal{N}} \Ya{}_{ab}\delta^{i}{}_{j} , 
\end{aligned}
\end{equation}
where the tensors $X_n$ are traceless. The remaining components $\tau_{ab}{}_i{}^{c}$ and $\tau_{ij}{}_a{}^{k}$ are already irreducible. We then find that for a compatible connection, $\tau$ has the vanishing irreducible parts
\begin{equation}
\label{eq:vanishing-parts}
\begin{aligned}
	&\alpha_{ij} = \beta_{ij} = -\tfrac23 \alpha_{ai} + \beta_{ai} = 0 , \\
	& \tau_{ab}{}_i{}^{c} = \tau_{ij}{}_a{}^{k} = 0 , \\
	& \Xb{}_{ij}{}^{k}{}_{l} = \Xd{}_{ai}{}^{j}{}_{k} = \Xf{}_{ab}{}^{i}{}_{j} = 0 ,\\
	& \tfrac16 \alpha_{ab}- \tfrac{1}{\mathcal{N}} \Ya{}_{ab} = \Yb{}_{ij} = 
		\Yc{}_{ia} + \tfrac{\mathcal{N}-1}{6} \alpha_{ai} = 0 ,
\end{aligned}
\end{equation}
and the respective conjugates, while all other components are allowed to be non-zero. The quantities in~\eqref{eq:vanishing-parts} thus parameterise the intrinsic torsion. These are then the explicit maps to the representations listed in~\ref{app:int-t}.

If we now allow $\Sigma$ to be a generic $\SU(8)$ connection, then looking back at the definitions~\eqref{eq:torsion-28-36-420} and~\eqref{eq:420-decomp} we see that requiring the parts~\eqref{eq:vanishing-parts} of its torsion to vanish is equivalent to fixing
\begin{equation}
\begin{aligned}
	\tau_{ij} = \tau_{ai} = 0 ,  \\
	\Sigma_{[ab}{}^i{}_{c]} = \Sigma_{[ab}{}^j{}_{i]} = \Sigma_{[ai}{}^k{}_{j]} = \Sigma_{[ij}{}^l{}_{k]} = 0 ,  \\
	\Sigma_{[ij}{}^a{}_{k]} = 0 , 
\end{aligned}
\end{equation}
as, for example,
\begin{equation}
\begin{aligned}
	\Sigma_{[ab}{}^j{}_{i]}  = \tau_{abi}{}^j + \tfrac12 \alpha_{[ab} \delta^j{}_{i]}
	= \Xf{}_{ab}{}^{j}{}_i - \tfrac{1}{\mathcal{N}} \Ya{}_{ab}\delta^{j}{}_{i}
		+ \tfrac16 \alpha_{ab} \delta^j{}_i .
\end{aligned}
\end{equation}
Thus we have recovered the conclusion of~\eqref{eq:int-torsion-red2}.



\end{document}